\documentclass[man,12pt,nolmodern,floatsintext]{apa7}
\usepackage{amsmath,amsthm,amssymb}
\usepackage{graphicx,float}
\usepackage[nodisplayskipstretch]{setspace}
\usepackage{enumitem}
\usepackage[american]{babel}
\usepackage{color}
\usepackage{verbatimbox}
\usepackage{caption}
\usepackage{mathtools}
\usepackage{multirow}
\usepackage{array}
\usepackage{xparse}
\usepackage{csquotes}
\usepackage{fancyhdr}
\usepackage{float}
\usepackage{url}
\usepackage[style=apa,backend=biber]{biblatex}
\usepackage{bm}
\usepackage{makecell}
\usepackage{pdfpages}

\restylefloat{table}
\restylefloat{figure}

\def\ex{\mathbb{E}}
\def\bX{\mathbf{X}}
\def\bY{\mathbf{Y}}
\def\bx{\boldsymbol{x}}
\def\by{\boldsymbol{y}}
\def\var{\mathrm{Var}}
\def\cov{\mathrm{Cov}}
\def\t{^\top}
\def\xxi{\boldsymbol{\xi}}

\definecolor{blue1}{RGB}{60,100,178}
\definecolor{red1}{RGB}{254,92,92}

\title{A New Fit Assessment Framework for Common Factor Models Using Generalized Residuals}
\shorttitle{Generalized Residuals}
\authorsnames{Youjin Sung, Youngjin Han, Yang Liu}
\authorsaffiliations{University of Maryland}

\authornote{Correspondence should be made to Youjin Sung at 1230B Benjamin Bldg, 3942 Campus Dr, University of Maryland, College Park, MD, 20742. Email: yjsung@umd.edu. The authors would like to thank Drs. Hongyang Zhao and Patricia Alexander from University of Maryland, College Park, for providing the empirical data.}

\abstract{Assessing fit in common factor models solely through the lens of mean and covariance structures, as is commonly done with conventional goodness-of-fit (GOF) assessments, may overlook critical aspects of misfit, potentially leading to misleading conclusions. To achieve more flexible fit assessment, we extend the theory of generalized residuals \parencite{haberman2013gr}, originally developed for models with categorical data, to encompass more general measurement models. Within this extended framework, we propose several fit test statistics designed to evaluate various parametric assumptions involved in common factor models. The examples include assessing the distributional assumptions of latent variables and the functional form assumptions of individual manifest variables. The performance of the proposed statistics is examined through simulation studies and an empirical data analysis. Our findings suggest that generalized residuals are promising tools for detecting misfit in measurement models, often masked when assessed by conventional GOF testing methods.} 

\begin{document}
                \maketitle

\section{1. Introduction}

In the fields of education and psychology, it is often of interest to study constructs that are not directly observable. These constructs cannot be measured directly and are often formulated as latent variables (LVs). Instruments are administered to collect observable indicators associated with these LVs, which are referred to as manifest variables (MVs). The LVs can then be inferred from the collected MVs by fitting a measurement model that specifies the relationship between the LVs and MVs.

A widely used LV measurement model is the common factor model \parencite[e.g.,][]{bollen1989, joreskog1969, kaplan2008, kline2023, lawley1971, thurstone1947}, particularly when analyzing continuous data. The common factor model requires MVs to be linearly dependent on LVs, aiming to attribute the dependencies among MVs to LVs. The model has been applied in a wide range of settings, such as testing theories about the number of LVs and the pattern of MV-LV dependency (i.e., confirmatory factor analysis) or learning these structures from data (i.e., exploratory factor analysis). Other common uses include examining individual differences in latent traits or investigating MV-level characteristics, similar to applications of item response theory \parencite[IRT; e.g.,][]{hambleton2013item}, but focusing on continuous MVs rather than categorical ones. Examples of such continuous responses include response times, often recorded in computer-based testing environments, and responses made by marking a placement along a line segment \parencite{minchen2018general}.

A crucial aspect of all these applications is the evaluation of model-data fit, as a poorly fitted model can invalidate subsequent inferences. For common factor models, numerous goodness-of-fit (GOF) assessment tools have been developed; among the most commonly reported (e.g., \cite{jackson2009reporting}) are the $\chi^2$ model fit test (i.e., likelihood ratio test) and several GOF indices, such as Comparative Fit Index (CFI; \cite{bentler1990}), Tucker-Lewis Index \parencite[TLI;][]{tucker1973}, Standardized Root Mean Squared Residual (SRMR; \cite{bentler1995}), and Root Mean Square Error of Approximation \parencite[RMSEA;][]{steiger1980}. If these fit diagnostics indicate good fit, researchers would generally be satisfied with the model and proceed with subsequent inferences.

However, conventional GOF statistics 
are primarily based on residual means and covariances (i.e., the differences between observed and fitted mean vectors and covariance matrices), which limits their ability to detect misfit occurring beyond the mean and covariance structures. It implies that even when conventional GOF diagnostics appear satisfactory, some masked misfit may still exist that is not captured by these measures. When the fit to mean and covariance structures is poor, researchers often turn to modification indices, which approximate changes in the model $\chi^2$ statistic in response to relaxing model constraints (e.g., allowing for cross-loadings or correlated residuals). Although these indices can help identify potential sources of misfit and guide model modifications, these adjustments remain confined to improving the recovery of mean and covariance structures.

In this paper, we emphasize that assessing GOF to mean and covariance structures only may overlook critical aspects of model-data disagreement, potentially resulting in misleading conclusions. For instance, in a common factor model with two LVs, the relation between the LVs is often of interest, typically inferred from the estimated correlation coefficient. When bivariate normality is assumed for the LVs, the correlation coefficient quantifies the direction and strength of their linear relation. If normality is violated, the correlation coefficient taken at face value may inaccurately represent the inter-LV association, as the two LVs may not even be linearly related. Such violations, however, cannot easily be detected by conventional methods. Another aspect that could potentially be overlooked is MV-level misfit. While residual means and variances can be inspected for individual MVs, this often provides little insight because the observed means and variances are often perfectly recovered in standard linear factor models.

To facilitate GOF assessment beyond mean and covariance structures, we focus on the theory of generalized residuals \parencite{haberman2013gr}, which was originally developed for and applied to categorical data organized in contingency tables. Informally, generalized residuals refer to the discrepancy between the sample average and the model-based population expectation of a specific summary quantity, which can be flexibly defined to evaluate various aspects of model misfit. This flexibility, coupled with the following benefits, makes generalized residuals a powerful tool for fit assessment. First, generalized residuals are asymptotically normal under the null hypothesis, enabling formal statistical tests. Second, they can be defined conditionally on LV values, allowing for localized fit assessments when combined with LV measurement models. When the latent dimensionality is low (e.g., one or two), graphical displays can also be conveniently generated conditionally on LV values, providing intuitive diagnostics of the nature of the misfit. Based on these features, the generalized residuals framework has been successfully applied in IRT with categorical data to evaluate specific parametric assumptions of the model, such as functional forms of item characteristic curves and the normality of LVs (\cite{haberman2019}; \cite{habermanetal2013}; \cite{monroe2021}; \cite{van2016}).

One contribution of our work is to extend the theory of generalized residuals
\parencite{haberman2013gr} to more general measurement models. We focus on GOF testing with linear normal factor analysis seeing its wide applications in measurement scenarios involving continuous MVs. As part of this extension, we introduce transformed residuals and establish their joint multivariate normality, which offers greater flexibility in constructing test statistics compared to the original framework. 
In addition, we propose summary statistics that evaluate multiple residuals simultaneously. 

The second contribution of our work is to propose three example test statistics that are designed to capture various aspects of model misfit in common factor models. These examples illustrate how our general framework can be adapted to specific fit assessment problems. The first example evaluates the fit of the LV density, while the second and third examples focus on MV-level fit assessment by examining the conditional mean and variance functions of each MV given the LVs. In all three examples, residuals are defined conditionally on LV values, allowing for both pointwise fit assessments at specific LV values and overall evaluations across multiple LV values. Given the flexibility of our framework, test statistics targeting other types of misfit can be explored in future research. The strength of our framework lies in its ability to identify issues that traditional GOF diagnostics may overlook, even when their results appear favorable. We demonstrate the utility of our method through simulations and an empirical data analysis using response time data. Ultimately, our proposed framework is expected to complement conventional GOF diagnostics for common factor models by enabling more nuanced assessments of model-data fit.

The remainder of the paper is organized as follows. In Section 2, we introduce the common factor model and present our extended theory of generalized residuals, along with example GOF test statistics. In Section 3, the performance of the proposed test statistics are evaluated by Monte Carlo studies. In Section 4, our GOF testing methods are illustrated with a real data example. The paper concludes with a discussion of the main findings, limitations, and future directions in Section 5.

\section{2. Theory}
 \subsection{2.1. Common Factor Model}

 \subsubsection{2.1.1. Latent Variable Measurement Model}
Let $Y_{ij} \in \mathcal{R}$ be the individual $i$'s response for manifest variable (MV) $j$, and $\bX_i=(X_{i1}, \dots, X_{id})^{\t} \in \mathcal{R}^d$ be the $d$-dimensional latent variable (LV) for individual $i$. Denote the conditional density of $Y_{ij}|\bX_{i}$ by $f_j(y_{ij}|\bx)$, where $y_{ij}$ and $\bx=(x_1, \dots, x_d)^{\t}$ indicate the realized values of $Y_{ij}$ and $\bX_i$, respectively. Let $\bY_i=(Y_{i1}, \dots , Y_{im})\t$ be a collection of $m$ MVs from the same individual $i$. It is typically assumed that $Y_{i1}, \dots , Y_{im}$ are conditionally independent given $\bX_i$, often referred to as the local independence assumption (\cite{McDonald1999}. pp. 255-257). The conditional density of $\bY_i=\by_i=(y_{i1}, \dots , y_{im})\t$ given $\bX_i=\bx$ can then be factorized as\footnote{In a more general context where the local independence assumption is relaxed, the product term on the right-hand side is replaced by the joint likelihood on the left-hand side in the following equations, with all other components remaining unchanged.}
\begin{equation}
    f(\by_i|\bx)=\prod_{j=1}^{m}f_j(y_{ij}|\boldsymbol{x}). 
\label{eq:cond_i}
\end{equation}
It follows that the marginal density/likelihood of $\bY_i=\by_i$ is obtained as the following integral
\begin{equation}
    f(\by_i)=\int \prod_{j=1}^{m}f_j(y_{ij}|\bx)\phi(\bx) d\bx,
\label{eq:mar.f}
\end{equation}
in which $\phi$ denotes the population density of $\bX_i$.

Pooling across a sample of $n$ independent and identically distributed (i.i.d.) random vectors of MVs, the sample log-likelihood can be expressed as 
\begin{equation}
   l_n(\boldsymbol{\xi}; \bY) = \sum_{i=1}^n \log f(\by_i; \boldsymbol{\xi}),
\label{eq:LL}
\end{equation}
in which $\bY=(\bY_1, \dots, \bY_n)^\top \in \mathcal{R}^{n \times m}$ denotes the data matrix, and the dependency on model parameters, denoted as $\boldsymbol{\xi} \in \mathcal{R}^q$, is now included in the notation. In parametric LV measurement models where the distribution of LVs and the conditional distribution of MVs given LVs are specified, $\boldsymbol{\xi}$ is often estimated by maximum likelihood (ML). ML estimation seeks for $\boldsymbol{\xi}$ that solves the following estimating equation 
\begin{equation}
 \nabla_{\boldsymbol{\xi}}l_n(\boldsymbol{\xi}; \bY)  = \mathbf{0},
\label{eq:est_eq}
\end{equation}
in which $\nabla_{\boldsymbol{\xi}}l_n(\boldsymbol{\xi}; \bY)$ is a $1 \times q$ vector of partial derivatives of $l_n(\boldsymbol{\xi}; \bY)$ with respect to $\boldsymbol{\xi}$. Provided that the Hessian matrix is negative definite, the solution to Equation \ref{eq:est_eq}, denoted $\hat{\boldsymbol{\xi}}$, corresponds to a local maximizer of the sample log-likelihood function under suitable regularity conditions. Given correct model specification, $\hat{\boldsymbol{\xi}}$ satisfies 
\begin{equation}
    \sqrt{n}(\hat{\boldsymbol{\xi}}-\boldsymbol{\xi}) \overset{d}{\to} \mathcal{N}\left(\mathbf{0},\bm{\mathcal{I}}^{-1}(\boldsymbol{\xi}) \right)
    \label{eq:info}
\end{equation}
as $n \to \infty$, where $\bm{\mathcal{I}}(\boldsymbol{\xi})=\ex [\nabla_{\boldsymbol{\xi}} \log f(\bY_i; \boldsymbol{\xi})^{\t} \nabla_{\boldsymbol{\xi}} \log f(\bY_i; \boldsymbol{\xi})]$ denotes the $q \times q$ per-observation Fisher information matrix. By Equation \ref{eq:info}, local identifiability of the model parameters is implicitly assumed.

\subsubsection{2.1.2. Common Factor Model}

In the current work, we focus on a specific LV measurement model for continuous MVs: the common factor model. We assume that $\bX_i \sim \mathcal{N}(\mathbf{0}, \boldsymbol{\Phi})$, with the population density of $\bX_i=\bx$ expressed by
\begin{align}
  \phi(\bx)=(2\pi)^{-\frac{d}{2}} \text{det}(\boldsymbol{\Phi})^{-\frac{1}{2}} \exp\left(-\frac{1}{2}\bx^{\t} \boldsymbol{\Phi}^{-1} \bx \right).
\label{eq:phix}
\end{align} 
We further assume that $Y_{ij}|\bX_i=\bx \sim \mathcal{N}(\nu_j+\boldsymbol{\lambda}_j^{\t} \bx, \theta_j)$ for $j=1,\dots,m$, where $\nu_j$ denotes an intercept, $\boldsymbol{\lambda}_j$ denotes a $d \times 1$ vector of factor loadings, and $\theta_j$ denotes an error variance. With this model specification, the conditional density $f_j(y_{ij}|\bx)$ is given by 
\begin{align}
   f_j(y_{ij}|\bx)=\frac{1}{\sqrt{2\pi \theta_j}}\exp\left(-\frac{1}{2\theta_j}(y_j-(\nu_j+\boldsymbol{\lambda}_j^{\t} \bx))^2\right).
\end{align}
The marginal likelihood $f(\by_i)$ can then be derived from Equation \ref{eq:mar.f}, resulting in a normal density function for $\mathcal{N}(\boldsymbol{\nu}, \boldsymbol{\Lambda} \boldsymbol{\Phi} \boldsymbol{\Lambda}^\top + \boldsymbol{\Theta})$: 
\begin{align}
  f(\by_i)=(2\pi)^{-\frac{m}{2}} \text{det}(\boldsymbol{\Lambda} \boldsymbol{\Phi} \boldsymbol{\Lambda}^\top + \boldsymbol{\Theta})^{-\frac{1}{2}} \exp\left(-\frac{1}{2}(\by_i - \boldsymbol{\nu})^{\t} (\boldsymbol{\Lambda} \boldsymbol{\Phi} \boldsymbol{\Lambda}^\top + \boldsymbol{\Theta})^{-1} (\by_i - \boldsymbol{\nu}) \right),
\label{eq:marginal}
\end{align} 
in which $\boldsymbol{\nu}=(\nu_1, \dots, \nu_m)^{\t}$ is a $m \times 1$ vector of intercepts, $\boldsymbol{\Lambda}=(\boldsymbol{\lambda}_1, \dots, \boldsymbol{\lambda}_m)^{\t}$ is a $m \times d$ matrix of factor loadings, and $\boldsymbol{\Theta} = \text{Diag}(\theta_1,\dots,\theta_m)$ is an $m \times m$ diagonal error covariance matrix.
The model parameters $\boldsymbol{\xi}$ now consist of the free elements in $\boldsymbol{\nu}$, $\boldsymbol{\Lambda}$, $\boldsymbol{\Phi}$, and $\boldsymbol{\Theta}$. The ML estimator $\hat{\boldsymbol{\xi}}$ can be obtained by solving Equation \ref{eq:est_eq} with the normal log-likelihood $l_n(\boldsymbol{\xi}; \bY)$. 

\subsubsection{2.1.3. Summary Quantity and Fit Assessment}

Under this common factor model, our goal now is to develop test statistics that can capture misfit beyond the mean and covariance structures. To achieve this, consider a summary quantity $H(\bY_i; \boldsymbol{\xi})$ for each individual $i$, which is a real continuous function allowed to depend on the MVs $\bY_i$ and the model parameters $\boldsymbol{\xi}$. Depending on the context, this function can also incorporate fixed LV values $\bx$. Using $H(\bY_i; \boldsymbol{\xi})$, we obtain its population expectation by
\begin{equation}
    \eta(\boldsymbol{\xi})=\ex[H(\bY_i; \boldsymbol{\xi})]=\int H( \by_i; \boldsymbol{\xi}) f(\by_i; \boldsymbol{\xi})d\by_i.
\label{eq:eta_single}
\end{equation}
An empirical estimator of Equation \ref{eq:eta_single} is the sample average
\begin{equation}
    \hat{\eta}(\boldsymbol{\xi})=\frac{1}{n}\sum_{i=1}^n H(\bY_i; \boldsymbol{\xi}).
\label{eq:etahat_single}
\end{equation}
The difference $\hat{\eta}(\boldsymbol{\xi})-\eta( \boldsymbol{\xi})$ then quantifies the discrepancy between the sample average and the model-implied population expectation, forming the basis for assessing model fit. Because model parameters are typically unknown in practice and must be estimated from data, we replace $\xxi$ by the ML estimator and rely on the residual $\hat{\eta}(\hat{\xxi})-\eta( \hat{\xxi})$ as the final basis for constructing fit statistics. 

When the same idea is applied to models for categorical data, the difference between sample and population averages of summary quantities is referred to as generalized residuals \parencite{haberman2013gr}. Generalized residuals are asymptotically normal under the correct model specification, which has enabled their wide application in constructing various fit statistics in IRT (e.g., \cite{haberman2019}; \cite{habermanetal2013}; \cite{monroe2021}; \cite{van2016}). 

In this paper, we extend the existing theory of generalized residuals \parencite{haberman2013gr} to establish asymptotic properties of transformed residuals under more general measurement models. Our proposed framework not only accommodates continuous MVs, but also enables flexible construction of diverse fit test statistics not considered in the original work on generalized residuals. In the following sections, we begin by presenting the most general theory in Section 2.2., construct two different types of test statistics in Section 2.3., and then narrow down to specific examples of fit assessment in Section 2.4.

\subsection{2.2. Residuals}

  \subsubsection{2.2.1. Asymptotic Normality of Residuals}

Consider a $k \times 1$ vector of summary quantities
\begin{equation}
\mathbf{H}(\bY_i; \boldsymbol{\xi})=(H_1(\bY_i; \boldsymbol{\xi}), H_2(\bY_i; \boldsymbol{\xi}), \dots, H_k(\bY_i; \boldsymbol{\xi}))^\top,
\label{eq:g.H}
\end{equation}
whose components are formed by different choices of summary quantities introduced in Section 2.1.3. Analogous to Equations \ref{eq:eta_single} and \ref{eq:etahat_single}, let $\boldsymbol{\eta}(\boldsymbol{\xi})$ and $\hat{\boldsymbol{\eta}}(\boldsymbol{\xi})$ denote the population expectation and sample average of $\mathbf{H}(\bY_i; \boldsymbol{\xi})$, respectively:
\begin{align}
\boldsymbol{\eta}(\boldsymbol{\xi}) &= \ex[\mathbf{H}(\bY_i; \boldsymbol{\xi})], \label{eq:eta.g}\\
    \hat{\boldsymbol{\eta}}(\boldsymbol{\xi}) &=\frac{1}{n}\sum_{i=1}^n \mathbf{H}(\bY_i; \boldsymbol{\xi}). \label{eq:etahat.g}
\end{align}
With the ML estimates for model parameters plugged in, the $k \times 1$ difference vector $\hat{\boldsymbol{\eta}}(\hat{\boldsymbol{\xi}})-\boldsymbol{\eta}(\hat{\boldsymbol{\xi}})$ quantifies the discrepancy between the data and the model, evaluated based on $H_\ell(\bY_i; \hat{\boldsymbol{\xi}})$ for $\ell=1,\dots,k$.

To further generalize, let $\boldsymbol{\varphi}: \mathcal{R}^{k} \to \mathcal{R}^{k'}$ be a twice continuously differentiable transformation function applied to both $\hat{\boldsymbol{\eta}}$ and $\boldsymbol{\eta}$. This transformation allows us to construct a $k' \times 1$ vector of \emph{transformed residuals}: 
\begin{equation}
   \boldsymbol{e}_{\boldsymbol{\varphi}}=\boldsymbol{\varphi}(\hat{\boldsymbol{\eta}}(\hat{\boldsymbol{\xi}}))-\boldsymbol{\varphi}(\boldsymbol{\eta}(\hat{\boldsymbol{\xi}})), 
\label{eq:trans.res}
\end{equation}
which evaluates model-data fit with respect to the population quantities of interest $\boldsymbol{\varphi}(\boldsymbol{\eta}(\xxi))$. This general formulation enables more flexible construction of statistics, including those involving ratios of summary quantities, as will be illustrated in Sections 2.4.2. and 2.4.3.

In the supplementary document, we establish the following asymptotic normality of the transformed residuals
\begin{align}
\sqrt{n}\boldsymbol{e}_{\boldsymbol{\varphi}}
\overset{d} \to \mathcal{N}\left(\mathbf{0},\nabla\boldsymbol{\varphi}(\boldsymbol{\eta}(\boldsymbol{\xi})) (-\mathbf{A}( \boldsymbol{\xi})\bm{\mathcal{I}}^{-1}(\boldsymbol{\xi})\mathbf{A}( \boldsymbol{\xi})^{\top} + \mathbf{\Sigma}_{\mathbf{H}}( \boldsymbol{\xi})) \nabla\boldsymbol{\varphi}(\boldsymbol{\eta}(\boldsymbol{\xi}))^\top \right),
\label{eq:ACM1}
\end{align}
in which $\nabla\boldsymbol{\varphi}$ stands for a $k' \times k$ Jacobian matrix of $\boldsymbol{\varphi}$, $\mathbf{A}(\boldsymbol{\xi})=\ex\left[\mathbf{H}(\bY_i; \boldsymbol{\xi}) \nabla_{\boldsymbol{\xi}} \log f(\bY_i; \boldsymbol{\xi}) \right]$, and $\mathbf{\Sigma}_{\mathbf{H}}(\boldsymbol{\xi})=\cov[\mathbf{H}( \bY_i; \boldsymbol{\xi})].$ 
In the special case where the identity transformation $\boldsymbol\varphi = \boldsymbol{\iota}$ is applied, with $\boldsymbol{\iota}(\mathbf{h})=\mathbf{h}$ for any $\mathbf{h}\in\mathcal{R}^{k}$, Equation \ref{eq:trans.res} reduces to a $k \times 1$ vector of original residuals, 
\begin{equation}
\boldsymbol{e}_{\boldsymbol{\iota}}=\hat{\boldsymbol{\eta}}(\hat{\boldsymbol{\xi}})-\boldsymbol{\eta}(\hat{\boldsymbol{\xi}}).
\label{eq:res.Ident}
\end{equation}
In this case, $\nabla\boldsymbol{\varphi}(\boldsymbol{\eta}(\boldsymbol{\xi}))$ in Equation \ref{eq:ACM1} simplifies to the $k \times k$ identity matrix. 

  \subsubsection{2.2.2. Estimation of the Asymptotic Covariance Matrix}
  
In practice, the asymptotic covariance matrix (ACM) of residuals (Equations \ref{eq:ACM1}) needs to be estimated from the data. When the population ACM is substituted with its consistent estimator, the asymptotic normality results remain unchanged by the Slustky's theorem \parencite[512]{bickel2015}. 

To obtain the consistent estimator of the ACM in Equation \ref{eq:ACM1}, we replace $\boldsymbol{\xi}$ with the ML estimator $\hat{\boldsymbol{\xi}}$ wherever they appear in the formula. %A caret "\^{}" is added to the notation of ACMs to highlight evaluation at the ML estimator $\hat{\boldsymbol{\xi}}$. 
For the population mean and (co)variance in
$\mathbf{A}(\hat{\boldsymbol{\xi}})$ and $\mathbf{\Sigma}_{\mathbf{H}}(\hat{\boldsymbol{\xi}})$, a straightforward approach is to use the sample mean and sample covariance matrix, as done by \textcite{habermanetal2013} and \textcite{monroe2021}.
However, our pilot simulation study revealed that this approach results in severely inflated Type I error rates with extreme LV values, particularly with small sample sizes. To address this issue, we suggest obtaining more stable estimates for $\mathbf{A}( \hat{\boldsymbol{\xi}})$ and $\mathbf{\Sigma}_{\mathbf{H}}( \hat{\boldsymbol{\xi}})$ by computing the Monte Carlo (MC) mean and covariance. This MC approach has been similarly used to approximate the expected Fisher information for IRT models (\cite{monroe2019estimation}). The steps are outlined as follows:

\begin{enumerate}
    \item Sample $\tilde{\by}_i, i=1,\dots,M$, from the marginal distribution of $\bY_i$ with density $f(\by_i;\hat{\boldsymbol{\xi}})$.
    \item Use $\tilde{\by}_i$ to calculate $\mathbf{H}( \tilde{\by}_i; \hat{\boldsymbol{\xi}})$ and $\nabla_{\boldsymbol{\xi}} \log f(\tilde{\by}_i; \hat{\boldsymbol{\xi}})$.
    \item Estimate $\mathbf{A}(\hat{\boldsymbol{\xi}})$ and $\mathbf{\Sigma}_{\mathbf{H}}(\hat{\boldsymbol{\xi}})$ by
\begin{align}
&\hat{\mathbf{A}}(\hat{\boldsymbol{\xi}})=\frac{1}{M}\sum_{i=1}^M \left[\mathbf{H}(\tilde{\by}_i; \hat{\boldsymbol{\xi}}) \nabla_{\boldsymbol{\xi}} \log f(\tilde{\by}_i; \hat{\boldsymbol{\xi}})\right],\\
&\hat{\mathbf{\Sigma}}_{\mathbf{H}}(\hat{\boldsymbol{\xi}})= \frac{1}{M-1}\sum_{i=1}^M \left[\left(\mathbf{H}(\tilde{\by}_i; \hat{\boldsymbol{\xi}}) - \bar{\mathbf{H}}( \hat{\boldsymbol{\xi}})\right)\left(\mathbf{H}(\tilde{\by}_i; \hat{\boldsymbol{\xi}}) - \bar{\mathbf{H}}(\hat{\boldsymbol{\xi}})\right)^\top\right].\label{eq:M}
\end{align}
\end{enumerate}
In Equation \ref{eq:M}, $\bar{\mathbf{H}}(\hat{\boldsymbol{\xi}})=\frac{1}{M}\sum_{i=1}^M \mathbf{H}(\tilde{\by}_i; \hat{\boldsymbol{\xi}})$. 
By the law of large numbers and Slutsky's theorem, $\hat{\mathbf{A}}(\hat{\boldsymbol{\xi}}) \overset{p}\to \mathbf{A}(\boldsymbol{\xi})$ and $\hat{\mathbf{\Sigma}}_{\mathbf{H}}(\hat{\boldsymbol{\xi}}) \overset{p}\to \mathbf{\Sigma}_{\mathbf{H}}(\boldsymbol{\xi})$. 

In the subsequent notation for ACMs, a caret ``\^{}'' is added to indicate that the ACMs are obtained using $\hat{\boldsymbol{\xi}}, \hat{\mathbf{A}},$ and $\hat{\mathbf{\Sigma}}_{\mathbf{H}}$.

\subsection{2.3. Test Statistics}

Based on the asymptotic normality of residuals established in Section 2.2.1., we develop two types of test statistics for assessing model misfit: (1) a $z$-statistic for testing a single residual, and (2) a $\chi^2$-statistic for testing multiple residuals simultaneously. 

   \subsubsection{2.3.1. $z$-statistic}

Let $e_{\boldsymbol{\varphi},r}$ be the $r$th component of $\boldsymbol{e}_{\boldsymbol{\varphi}}$ in Equation \ref{eq:trans.res} ($r=1, \dots, k'$).
Similarly, let $\hat{\sigma}_{\boldsymbol{\varphi},r}$ be the square root of the $r$th diagonal element of $\hat{\mathbf{\Sigma}}_{\boldsymbol{\varphi}}$, where the shorthand notation $\mathbf{\Sigma}_{\boldsymbol{\varphi}}$ is introduced to denote the ACM in Equation \ref{eq:ACM1}:
\begin{equation} 
   \mathbf{\Sigma}_{\boldsymbol{\varphi}}(\boldsymbol{\xi})=\nabla\boldsymbol{\varphi}(\boldsymbol{\eta}( \boldsymbol{\xi})) (-\mathbf{A}( \boldsymbol{\xi})\bm{\mathcal{I}}^{-1}(\boldsymbol{\xi})\mathbf{A}( \boldsymbol{\xi})^{\top} + \mathbf{\Sigma}_{\mathbf{H}}(\boldsymbol{\xi})) \nabla\boldsymbol{\varphi}(\boldsymbol{\eta}(\boldsymbol{\xi}))^\top.
\end{equation}
Using a single residual $e_{\boldsymbol{\varphi},r}$, we standardize it to construct the $z$-statistic:
\begin{equation}
z=\frac{e_{\boldsymbol{\varphi},r}}{\hat{\sigma}_{\boldsymbol{\varphi},r} / \sqrt{n}}.
\label{eq:z}
\end{equation}
Under the null hypothesis that the model is correctly specified,
the $z$-statistic converges in distribution to $\mathcal{N}(0, 1)$. Consequently, values of $|z|$ > 1.96 suggest significant misfit at $\alpha=0.05$ based on $e_{\boldsymbol{\varphi},r}$. 

  \subsubsection{2.3.2. $\chi^2$-statistic}
  
Building on the joint asymptotic multivariate normality of $\boldsymbol{e}_{\boldsymbol{\varphi}}$ established in Equation \ref{eq:ACM1}, we construct a quadratic form statistic $n\boldsymbol{e}_{\boldsymbol{\varphi}}^\top \mathbf{W} \boldsymbol{e}_{\boldsymbol{\varphi}}$, where $\mathbf{W}$ is a $Q \times Q$ symmetric weight matrix. This statistic provides a summary of model misfit based on multiple residuals. Such summary information can help practitioners make informed judgments about the overall model misfit of interest. In the remainder of this section, we first review some existing approaches for constructing quadratic form statistics based on normal variables, followed by our newly proposed approach.

When the model is correctly specified, it is known that a quadratic form statistic of normal variates is distributed as a mixture of independent $\chi_1^2$ random variables \parencite{box1954}. The correct $p$-value for this mixture-$\chi^2$ distribution can be computed, for example, by using the inversion formula given in \textcite{imhof1961} or by adjusting the statistic based on its mean and variance, approximating a $\chi^2$ distribution \parencite{satorra1994, satterthwaite1946}. However, calculating $p$-values under a $\chi^2$-mixture reference can often be computationally challenging.

To simplify the $p$-value calculations, the weight matrix $\mathbf{W}$ can be chosen such that the resulting quadratic form statistic is asymptotically distributed as $\chi^2$. Specifically, this condition is met when   
\begin{equation} \mathbf{\Sigma}_{\boldsymbol{\varphi}}\mathbf{W}\mathbf{\Sigma}_{\boldsymbol{\varphi}}\mathbf{W}\mathbf{\Sigma}_{\boldsymbol{\varphi}}=\mathbf{\Sigma}_{\boldsymbol{\varphi}}\mathbf{W}\mathbf{\Sigma}_{\boldsymbol{\varphi}}
\label{eq:chi.cond}
\end{equation}
and $tr(\mathbf{W}\mathbf{\Sigma}_{\boldsymbol{\varphi}})=s$, where $s$ is the degrees of freedom of the resulting $\chi^2$ distribution (\cite{schott2016}, Theorem 11.11). 
A common way to satisfy these conditions is to use the Moore-Penrose pseudoinverse $\boldsymbol{\Sigma}_{\boldsymbol{\varphi}}^{+}$ of $\boldsymbol{\Sigma}_{\boldsymbol{\varphi}}$ \parencite[36]{magnus1999} as the weight matrix. This approach has been employed in previous studies to construct quadratic form fit statistics \parencite[e.g.,][]{liu2014, maydeuliu2015, liu2019, reiser1996}. However, the use of $\boldsymbol{\Sigma}_{\boldsymbol{\varphi}}^{+}$ can introduce numerical instability, as the rank of $\mathbf{\Sigma}_{\boldsymbol{\varphi}}$ needs to be estimated by counting non-zero eigenvalues in its sample estimate. This procedure relies on an arbitrarily selected tolerance for identifying zero eigenvalues, which may lead to inconsistent performance of the resulting test statistic \parencite{maydeu2008}.

To address this issue, we propose defining the weight matrix as 
\begin{equation}   \mathbf{W}=\mathbf{\Sigma}_{\boldsymbol{\varphi}}^{-}=\mathbf{U} \mathbf{\Omega}^{-} \mathbf{U}^\top,
\label{eq:w}
\end{equation}
in which $\mathbf{U}$ is the $Q \times Q$ matrix of eigenvectors from the eigendecomposition of $\mathbf{\Sigma}_{\boldsymbol{\varphi}}$ (i.e., $\mathbf{\Sigma}_{\boldsymbol{\varphi}}=\mathbf{U} \mathbf{\Omega} \mathbf{U}^\top$), $\mathbf{\Omega}$ is the diagonal matrix of eigenvalues, and    $\mathbf{\Omega}^{-}=\text{Diag}(\omega_1^{-1}, \dots, \omega_s^{-1}, 0, \dots, 0)$. Here, $\mathbf{\Omega}^{-}$ stands for the inverse of $\mathbf{\Omega}$ only with the largest $s$ eigenvalues, $\omega_1, \dots, \omega_s$, where $s \leq \text{rank}(\mathbf{\Sigma}_{\boldsymbol{\varphi}})$. The proposed weight matrix $\mathbf{\Sigma}_{\boldsymbol{\varphi}}^{-}$ generalizes the pseudoinverse $\boldsymbol{\Sigma}_{\boldsymbol{\varphi}}^{+}$, subsuming it as a special case when $s=\text{rank}(\mathbf{\Sigma}_{\boldsymbol{\varphi}})$. For $s \leq \text{rank}(\mathbf{\Sigma}_{\boldsymbol{\varphi}})$, $\mathbf{\Sigma}_{\boldsymbol{\varphi}}$ serves as a generalized inverse of $\mathbf{\Sigma}_{\boldsymbol{\varphi}}^{-}$, such that $\mathbf{\Sigma}_{\boldsymbol{\varphi}}^{-} \mathbf{\Sigma}_{\boldsymbol{\varphi}} \mathbf{\Sigma}_{\boldsymbol{\varphi}}^{-}=\mathbf{\Sigma}_{\boldsymbol{\varphi}}^{-}$ holds.
With this formulation, the conditions in Equation \ref{eq:chi.cond} and $tr(\mathbf{\Sigma}_{\boldsymbol{\varphi}}^{-}\mathbf{\Sigma}_{\boldsymbol{\varphi}})=s$ are satisfied.

Accordingly, we define our $\chi^2$-statistic using $\hat{\mathbf{W}}=\mathbf{\hat{\Sigma}}_{\boldsymbol{\varphi}}^{-}$ as
\begin{equation} 
 T=n\boldsymbol{e}_{\boldsymbol{\varphi}}^\top \hat{\mathbf{\Sigma}}_{\boldsymbol{\varphi}}^{-} \boldsymbol{e}_{\boldsymbol{\varphi}},
\label{eq:summary}
\end{equation}
such that $T$ asymptotically follows a $\chi^2$ distribution with $s$ degrees of freedom ($\chi^2_s)$ under correct model specification. Since this asymptotic property holds for any $1 \leq s \leq \text{rank}(\mathbf{\Sigma}_{\boldsymbol{\varphi}})$, a sufficiently small value, such as $s=1$, can be chosen as the reference degrees of freedom to circumvent the numerical instability issue. 

With the choice of $s=1$, the reference distribution becomes $\chi^2_1$, where $T > 3.84$ indicates significant overall misfit at $\alpha=0.05$ based on multiple residuals in $\boldsymbol{e}_{\boldsymbol{\varphi}}$.
The performance of the proposed $\chi^2$-statistic with $s=1$ will be evaluated through our simulation studies in Section 3.

\subsection{2.4. Examples}

In this section, we provide three examples to illustrate how our proposed framework can be adapted to assess specific model assumptions in common factor models. The first example (Section 2.4.1.) focuses on assessing the distributional assumption of LVs. The second and third examples (Sections 2.4.2. and 2.4.3.) address testing MV-level linearity and homoscedasticity by evaluating each MV's mean and variance functions conditional on the LVs. Note that the key connection from our general framework to practical model fit assessment lies in the careful design of the functions $\mathbf{H}$ and the selection of the transformation function $\boldsymbol{\varphi}$. Although we focus on only three examples here, further variations are open for future research.

 \subsubsection{2.4.1. Testing Normality of LV Density}

%Let $\bx_1, \dots, \bx_Q$ denote the $d$-dimensional grids of LV values across $Q$ points, for which we aim to assess the misfit in LV density. For a single grid, $\bx_\ell (\ell=1, \dots, Q)$, define 
Let $\bx_1, \dots, \bx_Q$ denote the $Q$ distinct values of the $d$-dimensional LV, for which we aim to assess the misfit in LV density. Given our goal of evaluating normality assumption of LVs, the dependency of summary quantities on LV values is now naturally introduced in our formulation, even when not explicitly expressed in the notation.

First, construct $\mathbf{H}$ in Equation \ref{eq:g.H} with $k=Q$ as
\begin{equation}
    \mathbf{H}(\bY_i; \boldsymbol{\xi})=(H_1(\bY_i; \boldsymbol{\xi}), H_2(\bY_i; \boldsymbol{\xi}), \dots, H_Q(\bY_i; \boldsymbol{\xi}))^\top,
\label{eq:g.H1}
\end{equation}
in which each component $H_\ell$ ($\ell=1, \dots, Q)$ represents the posterior density of $\bx_\ell$ given $\bY_i$:
\begin{equation}
    H_\ell(\bY_i; \boldsymbol{\xi})=f(\bx_\ell|\bY_i; \boldsymbol{\xi}).
\label{eq:H_ex1}
\end{equation}
With this choice of $H_\ell$, its population expectation, denoted by $\eta_\ell$, corresponds to the LV density $\phi(\bx_\ell; \boldsymbol{\xi})$ defined in Equation \ref{eq:phix}:
\begin{align}
   \eta_\ell(\boldsymbol{\xi}) &= \ex [H_\ell(\bY_i; \boldsymbol{\xi})]=\phi(\bx_\ell; \boldsymbol{\xi}).\label{eq:eta_ex1}
\end{align}
Equation \ref{eq:eta_ex1} follows from Bayes' theorem,
\begin{equation}
   f(\bx_\ell|\bY_i; \boldsymbol{\xi})=\frac{\phi(\bx_\ell; \boldsymbol{\xi})f(\bY_{i}|\bx_\ell; \boldsymbol{\xi})}{f(\bY_i; \boldsymbol{\xi})},
\label{eq:bayes}
\end{equation}
along with Equation \ref{eq:eta_single}.
The empirical estimator of $\eta_\ell$, denoted by $\hat{\eta}_\ell$, is then constructed as 
\begin{align}
      \hat{\eta}_\ell(\boldsymbol{\xi}) &= \frac{1}{n}\sum_{i=1}^n H_\ell(\bY_i; \boldsymbol{\xi})=\frac{1}{n}\sum_{i=1}^n f(\bx_\ell|\bY_i; \boldsymbol{\xi}),\label{eq:etahatH}
\end{align}
representing the sample mean of the posterior densities.
Now in the vector form, the difference between $\hat{\boldsymbol{\eta}}$ and $\boldsymbol{\eta}$ (Equations \ref{eq:etahat.g} and \ref{eq:eta.g}) collectively represents
the discrepancies between the population LV density and its sample estimate on $\bx_1, \dots, \bx_Q$.
Thus, with $\boldsymbol{\varphi}=\boldsymbol{\iota}$ being the identity transformation (see Equation \ref{eq:res.Ident}), the resulting $Q \times 1$ vector of residuals,
\begin{equation}
\boldsymbol{e}_{\boldsymbol{\iota}}=\hat{\boldsymbol{\eta}}(\hat{\boldsymbol{\xi}})-\boldsymbol{\eta}(\hat{\boldsymbol{\xi}}),
\label{eq:res.ex1}
\end{equation}
reflects the nonnormality of the LVs.

With this formulation, we construct the $z$-statistic from Equation \ref{eq:z} based on each single residual $e_{\boldsymbol{\iota}, \ell}=\hat{\eta}_\ell(\hat{\boldsymbol{\xi}})-\eta_\ell(\hat{\boldsymbol{\xi}})$, which captures the misfit in LV density at $\bx_\ell$. In addition, we construct the $\chi^2$-statistic from Equation \ref{eq:summary} based on $\boldsymbol{e}_{\boldsymbol{\iota}}$, which reflects the overall misfit in LV density across $\bx_1, \dots, \bx_Q$. Given the nature of fit assessment in our example, we will use the terms \textit{pointwise} statistic and \textit{summary} statistic interchangeably with the $z$-statistic and the $\chi^2$-statistic, respectively. Also, we denote these statistics developed for this LV density example as $z_1(\bx_\ell)$ and $T_1$, respectively, for use in later sections. The pointwise statistic $z_1$ coincides with the one proposed in the context of IRT for detecting misfit in the LV density under unidimensional IRT models \parencite{monroe2021}.

  \subsubsection{2.4.2. Testing MV-level Linearity}
On the same grid of LV values $\bx_1, \dots, \bx_Q$, 
construct $\mathbf{H}$ in Equation \ref{eq:g.H} with $k=2Q$ as follows:
\begin{equation}
   \mathbf{H}(\bY_i; \boldsymbol{\xi})=(H_{1}(\bY_i; \boldsymbol{\xi}), \dots, H_{Q}(\bY_i; \boldsymbol{\xi}), H_{Q+1}(\bY_i; \boldsymbol{\xi}), \dots, H_{2Q}(\bY_i; \boldsymbol{\xi}))^\top,
\label{eq:H_ratio}
\end{equation}
in which
\begin{align}
   &H_{\ell}(\bY_i; \boldsymbol{\xi})=Y_{ij} f(\bx_\ell|\bY_i; \xxi),  \label{eq:H1}\\
    &H_{Q+\ell}(\bY_i; \boldsymbol{\xi})=f(\bx_\ell|\bY_i; \xxi),\label{eq:H2}
\end{align}
for $\ell=1, \dots, Q$.
 
Following the construction of $\hat{\boldsymbol{\eta}}$ and $\boldsymbol{\eta}$ as in Equations \ref{eq:etahat.g} and \ref{eq:eta.g}, define the $Q \times 1$ vector of transformed residuals (i.e., $k'=Q$) as
\begin{equation}
   \boldsymbol{e}_{\boldsymbol{\varphi}}=\boldsymbol{\varphi}(\hat{\boldsymbol{\eta}}(\hat{\boldsymbol{\xi}}))-\boldsymbol{\varphi}(\boldsymbol{\eta}(\hat{\boldsymbol{\xi}})),
\label{eq:trans.res_ex2}
\end{equation}
in which a transformation function $\boldsymbol{\varphi}: \mathcal{R}^{2Q} \to \mathcal{R}^Q$ is applied to produce ratios:
\begin{equation}
\boldsymbol{\varphi}(\boldsymbol{\gamma})=\left(\frac{\gamma_1}{\gamma_{Q+1}}, \dots, \frac{\gamma_Q}{\gamma_{2Q}}\right)^\top
\label{eq:varphi}
\end{equation}
with $\boldsymbol{\gamma}=(\gamma_1,\dots,\gamma_Q,\gamma_{Q+1},\dots,\gamma_{2Q})^\top$.
With this setup, the residuals in Equation \ref{eq:trans.res_ex2} represent the misfit in the mean response for the $j$th MV, conditional on $\bx_1,\dots,\bx_Q$.

To elaborate, the expectations of $H_{\ell}$ and $H_{Q+\ell}$ are obtained by
\begin{align}
   &\eta_{\ell}(\boldsymbol{\xi}) = \ex [H_{\ell}(\bY_i; \boldsymbol{\xi})]=\phi(\bx_\ell; \boldsymbol{\xi})\ex(Y_{ij} | \bx_\ell; \xxi),\\
   &\eta_{Q+\ell}(\boldsymbol{\xi}) = \ex [H_{Q+\ell}(\bY_i; \boldsymbol{\xi})]=\phi(\bx_\ell; \boldsymbol{\xi})
   \label{eq:eta_2l}
\end{align}
from Equations \ref{eq:eta_single} and \ref{eq:bayes}, with corresponding empirical estimators
\begin{align}
      &\hat{\eta}_{\ell}(\boldsymbol{\xi}) = \frac{1}{n}\sum_{i=1}^n H_{\ell}(\bY_i; \boldsymbol{\xi})=\frac{1}{n}\sum_{i=1}^n Y_{ij}f(\bx_\ell|\bY_i; \boldsymbol{\xi}),\\
      &\hat{\eta}_{Q+\ell}(\boldsymbol{\xi}) = \frac{1}{n}\sum_{i=1}^n H_{Q+\ell}(\bY_i; \xxi)=\frac{1}{n}\sum_{i=1}^n f(\bx_\ell|\bY_i; \xxi).
      \label{eq:eta_2lhat}
\end{align}
After applying the ratio transformation function in Equation \ref{eq:varphi} to the vector form counterparts $\hat{\boldsymbol{\eta}}$ and $\boldsymbol{\eta}$, the $\ell$th component of $\boldsymbol{e}_{\boldsymbol{\varphi}}$ in Equation \ref{eq:trans.res_ex2} becomes 
\begin{equation}
    e_{\boldsymbol{\varphi}, \ell} = 
 \frac{\hat{\eta}_{\ell}(\hat{\xxi})}{\hat{\eta}_{Q+\ell}(\hat{\xxi})} - \frac{\eta_{\ell}(\hat{\xxi})}{\eta_{Q+\ell}(\hat{\xxi})} = 
    \frac{\sum_{i=1}^n Y_{ij}f(\bx_\ell|\bY_i; \hat{\xxi})}{\sum_{i=1}^n f(\bx_\ell|\bY_i; \hat{\xxi})} - \ex(Y_{ij} | \bx_\ell; \hat{\xxi}).
\label{eq:ratio}
\end{equation}
%[\boldsymbol{\varphi}(\hat{\boldsymbol{\eta}}({\hat\xxi})]_\ell - [\boldsymbol{\varphi}(\boldsymbol{\eta}({\hat\xxi})]_\ell 

Note that the conditional expectation in Equation \ref{eq:ratio}, $\ex(Y_{ij} | \bx_\ell; \hat{\xxi})$, is known as the item characteristic curve under a unidimensional IRT model with dichotomous response variables. Based on this, similar form of residuals has been proposed to assess item-level fit in IRT models (e.g., \cite{habermanetal2013}). In the context of the common factor model, the conditional expectation is simply a linear function of $\bx_\ell$: 
\begin{equation}
   \ex(Y_{ij} | \bx_\ell; \hat{\xxi})=\hat{\nu}_j+\boldsymbol{\hat{\lambda}}_j^{\t} \bx_\ell.
\label{eq:cond_exp}
\end{equation}
Therefore, the residual $e_{\boldsymbol{\varphi}, \ell}$ in Equation \ref{eq:ratio} serves as a basis for assessing this linearity assumption for the $j$th MV.

With this background, the pointwise statistic is now constructed from Equation \ref{eq:z} to assess the MV-level linearity at $\bx_\ell$, while the summary statistic is constructed from Equation \ref{eq:summary} to provide an overall assessment across multiple LV grids $\bx_1, \dots, \bx_Q$. We refer to these statistics as $z_2(\bx_\ell)$ and $T_2$, respectively.

Before moving on to the next example, it is worth noting that alternative ways to formulate residuals targeting the same population quantity may exist. For instance, with the target population quantity  $\ex(Y_{ij} | \bx_\ell; \xxi)$ as in this example, we can keep the formulation of $\mathbf{H}$ as in Equation \ref{eq:g.H1}, with each component defined as
\begin{equation}
    H_\ell(\bY_i; \boldsymbol{\xi})=\frac{Y_{ij}f(\bY_{i}|\bx_\ell; \xxi)}{f(\bY_i; \xxi)}.
\label{eq:H_ex2}
\end{equation}
With this choice of $H_\ell$, its population expectation $\eta_\ell$ directly becomes $\ex(Y_{ij} | \bx_\ell; \xxi)$, eliminating the need for a transformation and thus simplifying the formulation of residuals. The corresponding empirical estimator is
\begin{equation}
        \hat{\eta}_\ell(\xxi) = \frac{1}{n}\sum_{i=1}^n \frac{Y_{ij}f(\bY_i|\bx_\ell; \xxi)}{f(\bY_i; \xxi)}=\frac{1}{\phi(\bx_\ell; \xxi)} \frac{1}{n} \sum_{i=1}^n Y_{ij}f(\bx_\ell|\bY_i; \xxi).
\label{eq:simple}
\end{equation}
We observed in pilot simulations that the performance of generalized residuals based on Equation \ref{eq:simple} is very sensitive to misspecification of the LV density $\phi(\bx_\ell; \xxi)$. In contrast, the ratio formulation we presented in Equation \ref{eq:ratio} also estimates the LV density in Equation \ref{eq:simple}, resulting in
a more targeted fit diagnostic. Further discussion on this will be provided in the supplementary document using a real data example.

  \subsubsection{2.4.3. Testing MV-level Homoscedasticity}

In this example, we retain the setup from Section 2.4.2., modifying only the first $Q$ $H_{\ell}(\bY_i; \boldsymbol{\xi})$ defined in Equation \ref{eq:H1} as follows:
\begin{align}
   H_{\ell}(\bY_i; \boldsymbol{\xi})&=[Y_{ij}-\ex(Y_{ij}|\bx_\ell; \xxi)]^2 f(\bx_\ell|\bY_i; \xxi). \label{eq:H1_2}
\end{align}
The expectation of $H_{\ell}$ then becomes
\begin{align}
   \eta_{\ell}(\xxi) &= \ex [H_{\ell}(\bY_i; \xxi)]=\phi(\bx_\ell; \xxi)\var(Y_{ij} | \bx_\ell; \xxi)
\end{align}
with the empirical estimator
\begin{align}
      \hat{\eta}_{\ell}(\boldsymbol{\xi}) =\frac{1}{n}\sum_{i=1}^n [Y_{ij}-\ex(Y_{ij}|\bx_\ell; \xxi)]^2 f(\bx_\ell|\bY_i; \boldsymbol{\xi}).        \label{eq:etahat.ex3}
\end{align}
The definitions of $H_{Q+\ell}, \eta_{Q+\ell},$ and $\hat\eta_{Q+\ell}$ remain unchanged, as specified in Equations \ref{eq:H2}, \ref{eq:eta_2l}, and \ref{eq:eta_2lhat}.
After applying the ratio transformation function in Equation \ref{eq:varphi} to the corresponding vector forms $\boldsymbol{\eta}$ and $\boldsymbol{\hat\eta}$, the $\ell$th component of the resulting $\boldsymbol{e}_{\boldsymbol{\varphi}}$ becomes 
\begin{equation}
    e_{\boldsymbol{\varphi}, \ell} = \frac{\hat{\eta}_{\ell}(\hat{\xxi})}{\hat{\eta}_{Q+\ell}(\hat{\xxi})} - \frac{\eta_{\ell}(\hat{\xxi})}{\eta_{Q+\ell}(\hat{\xxi})} = 
    \frac{\sum_{i=1}^n [Y_{ij}-\ex(Y_{ij}|\bx_\ell;\hat{\xxi})]^2 f(\bx_\ell|\bY_i; \hat{\xxi})}{\sum_{i=1}^n f(\bx_\ell|\bY_i; \hat{\xxi})} - \var(Y_{ij} | \bx_\ell; \hat{\xxi}).
\label{eq:ratio2}
\end{equation}
Under the common factor model, the conditional variance of the $j$th MV in Equation \ref{eq:ratio2} is a constant (i.e., 
$\var(Y_{ij} | \bx_\ell; \hat{\xxi})=\hat{\theta}_j$) and does not depend on LVs $\bx_\ell$. Therefore, this residual can be used to assess the constant variance assumption for the $j$th MV. Similar to the previous examples, the pointwise statistic is constructed from Equation \ref{eq:z} to assess the MV-level homoscedasticity at $\bx_\ell$, while the summary statistic is constructed from Equation \ref{eq:summary} to summarize results across multiple LV grids $\bx_1, \dots, \bx_Q$. We refer to these statistics as $z_3(\bx_\ell)$ and $T_3$, respectively.

In summary, we have discussed three examples of fit assessments that can be developed within our framework. We constructed pointwise statistics, 
$z_1(\bx_\ell), z_2(\bx_\ell),$ and $z_3(\bx_\ell)$ ($\ell=1, \dots, Q$), and summary statistics, $T_1, T_2$ and $T_3$, optimized for testing the normality of LV density, MV-level linearity of the mean function, and MV-level homoscedasticity of the variance function, respectively. Table \ref{tb:example} provides a summary of the three examples discussed so far. 

\begin{table}[h]
\caption{Summary of components for constructing the three examples discussed in Section 2.4. For brevity, dependency on model parameters $\xxi$ is omitted. In the first column of the second row, $[\boldsymbol{\varphi}(\boldsymbol{\eta})]_\ell$ denotes the $\ell$th component of the transformed $\boldsymbol{\eta}$, representing the population quantity of interest. In these examples, the index $\ell$ distinguishes the LV values under evaluation, where $\ell = 1, \dots, Q$.}
\label{tb:example}
\centering
\begin{tabular}{c c c c}  
\hline
   Sources of misfit & LV density & \makecell{MV-level \\ linearity} & \makecell{MV-level \\ homoscedasticity} \\
\hline
   \makecell{Target population \\ quantity $[\boldsymbol{\varphi}(\boldsymbol{\eta})]_\ell$} & $\phi(\bx_\ell)$ & $\ex(Y_{ij}|\bx_\ell)$ & $\var(Y_{ij}|\bx_\ell)$ \\
   Choices of $H_\ell$ & $\makecell{H_\ell=f(\bx_\ell|\bY_i)}$ & 
   \makecell{$H_{\ell}=Y_{ij} f(\bx_\ell|\bY_i)$ \\ $H_{Q+\ell}=f(\bx_\ell|\bY_i)$} & 
   \makecell{$H_{\ell}=[Y_{ij}-\ex(Y_{ij}|\bx_\ell)]^2 f(\bx_\ell|\bY_i)$ \\ $H_{Q+\ell}=f(\bx_\ell|\bY_i)$} \\
   Transformation $\boldsymbol{\varphi}$ & Identity & Ratio & Ratio \\
   Pointwise statistic & $z_1(\bx_\ell)$ & $z_2(\bx_\ell)$ & $z_3(\bx_\ell)$ \\
   Summary statistic  & $T_1$ & $T_2$ & $T_3$ \\
\hline
\end{tabular}
\end{table}

\section{3. Simulation Study}

To evaluate the finite sample behaviors of the proposed test statistics under our framework, two simulation studies were conducted. Study 1 (Section 3.1.) evaluates the performance of $z_1(\bx_\ell)$ and $T_1$, while Study 2 (Section 3.2.) evaluates the performance of $z_2(\bx_\ell), T_2$ and $z_3(\bx_\ell), T_3$. Within each study, we manipulate two factors: (1) sample sizes ($n=100, 500, 1000$), representing small, moderate, and large samples, and (2) the presence or absence of model misspecification. For each simulation condition, pointwise $z$-tests at $\bx_1, \dots, \bx_Q$ and an overall $\chi^2$-test are performed with one simulated dataset, which is replicated 500 times to obtain empirical rejection rates under each condition.

In the following subsections addressing each study, we first describe the simulation setup, including the simulation conditions and data-generating models, followed by the detailed procedures for calculating the test statistics (Sections 3.1.1. and 3.2.1.). We then examine their performance using empirical rejection rates as the evaluation criteria (Sections 3.1.2. and 3.2.2.). As no convergence issues were found in either study, empirical rejection rates were calculated as the proportion of rejected cases among the 500 replications. At the end of each study, results from some conventional fit diagnostics are also presented for comparison. All computations were performed in R \parencite{R}. Example code is included in the Online Supplementary Materials.

\subsection{3.1. Simulation Study 1}
 
 \subsubsection{3.1.1. Simulation Setup}
 
As a special case of the common factor model presented in Section 2.1.2., we considered a two-dimensional independent-cluster model with LVs: $\bX_i=(X_{i1}, X_{i2})^{\t}$. The number of MVs was set to twenty (i.e., $m=20$), with the first ten MVs loaded only on $X_{i1}$, and the remaining ten only on $X_{i2}$. The two LVs were assumed to be correlated with each other. 

\begin{figure}[t]
    \centering
    \includegraphics[width=130mm]{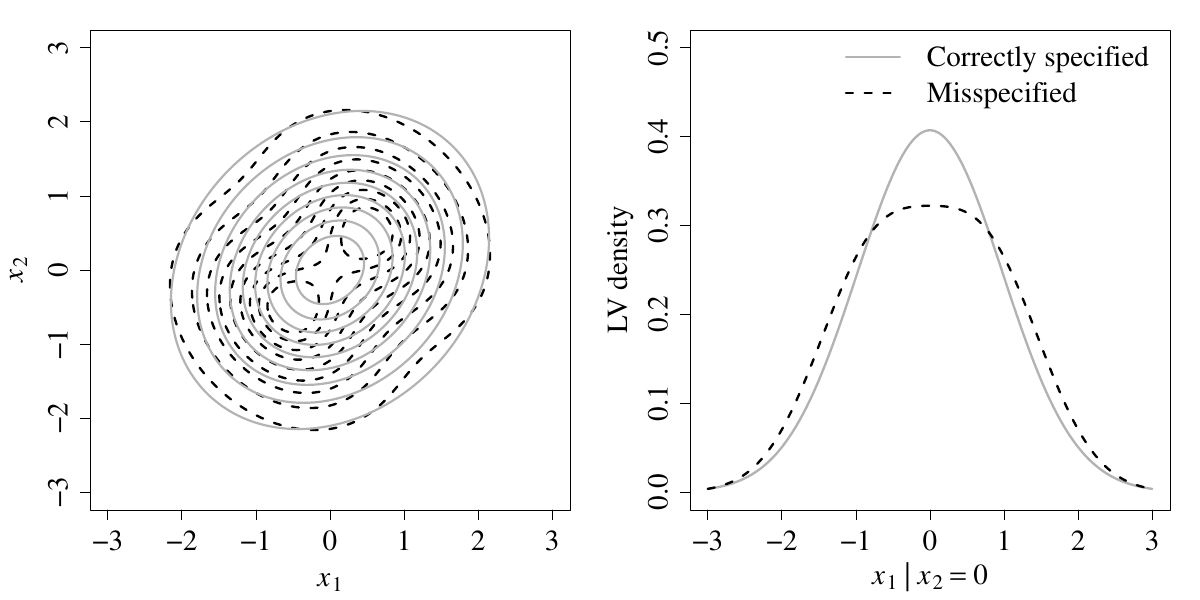}
    \caption{Left panel: Contour plot of the data-generating LV densities under correctly specified (grey solid lines) and misspecified (black dashed lines) conditions. Right panel: Conditional density of $X_{i1}=x_1$ given $X_{i2}=x_2=0$ under the same conditions. The conditional density of $X_{i2}=x_2$ given $X_{i1}=x_1=0$ is identical to that shown in the right panel.}
    \label{fig:LVD}
\end{figure}

For data generation, two different LV densities were specified to represent correctly specified and misspecified conditions. For the correctly specified condition, $\bX_i, i=1, \dots, n,$ were generated from the standard bivariate normal distribution with a correlation coefficient of 0.2. For the misspecified condition, we simulated the LVs from a mixture of two bivariate normal distributions: $\bX_i \sim 0.5\mathcal{N}(\boldsymbol{\mu}^{(1)}, \mathbf{\Phi}^{(1)})+0.5\mathcal{N}(\boldsymbol{\mu}^{(2)}, \mathbf{\Phi}^{(2)}),$ 
in which $\boldsymbol{\mu}^{(1)}=(-0.6, -0.6)^\top$, $\boldsymbol{\mu}^{(2)}=(0.6, 0.6)^\top$, and 
\begin{equation}
\mathbf{\Phi}^{(1)}=\mathbf{\Phi}^{(2)}=\begin{bmatrix}
    0.8^2 & -0.16  \\
    -0.16 & 0.8^2
  \end{bmatrix}.
\end{equation}
With this setup, the LV distributions in both conditions have means of zero and the covariance matrix of
\begin{equation}
\mathbf{\Phi}=\cov(\mathbf{X}_i)=
\begin{bmatrix}
    1 & 0.2  \\
    0.2 & 1
  \end{bmatrix}.
  \label{eq:phi}
\end{equation} 
The left panel of Figure \ref{fig:LVD} depicts the contour plot of the two data-generating LV densities, with one superimposed on the other. The right panel displays the conditional density of $X_{i1}=x_1$ given the mean of $X_{i2}$,  which is zero in this case. The right panel will be revisited in the discussion of results in Section 3.1.2.

In both correctly and incorrectly specified conditions, other data-generating parameters were set equal. In particular, MVs were assumed to have zero intercepts and exhibit an independent-cluster factor structure: $\boldsymbol{\nu}=\mathbf{0}$ and 
\begin{equation}
   \mathbf{\Lambda}=
   \begin{bmatrix}
       \boldsymbol{a}_1 & \mathbf{0}_{10} \\
        \mathbf{0}_{10} & \boldsymbol{a}_2\\
   \end{bmatrix},
\end{equation}
in which $\boldsymbol{a}_1=\boldsymbol{a}_2=(\sqrt{0.3},\sqrt{0.5},\sqrt{0.7},\sqrt{0.3},\sqrt{0.5},\sqrt{0.7},\sqrt{0.3},\sqrt{0.5},\sqrt{0.7},\sqrt{0.3})^\top$ and $\mathbf{0}_{10}$ indicates a $10 \times 1$ vector of zeros. Error variances $\theta_j, j=1, \cdots, 20,$ were then determined by diagonal elements of the square matrix $\mathbf{I}-\boldsymbol{\Lambda} \boldsymbol{\Phi} \boldsymbol{\Lambda}^\top$. This model specification implies that the communalities alternate among 0.3, 0.5, and 0.7, covering a range of values typically observed in practice. A value of 0.3 represents low communality, while 0.7 represents high communality (\cite{maccallum1999sample}). The value of 0.5 was chosen as an intermediate point to reflect moderate communality.

Upon generating data using the specified parameter values, the proposed tests for assessing the LV density fit were conducted using the following procedures: (1) Estimate the model parameters $\xxi$ by ML to get $\hat\xxi$, (2) construct $\boldsymbol{\hat{\eta}}(\hat\xxi)$ and $\boldsymbol{\eta}(\hat\xxi)$ and calculate residuals by $\boldsymbol{e}_{\boldsymbol{\iota}}=\boldsymbol{\hat{\eta}}(\hat\xxi)-\boldsymbol{\eta}(\hat\xxi)$, (3) estimate the ACM of $\boldsymbol{e}_{\boldsymbol{\iota}}$, (4) compute $z_1(\bx_\ell)$ for $\ell=1,\dots, Q$ and $T_1$ to conduct pointwise $z$-tests and the overall $\chi^2$-test.

For parameter estimation in Step 1, the {\tt lavaan} package version 0.6-14 \parencite{lavaan} was used with the default option of ML estimation with the normal likelihood. Mean structure was added to the model ({\tt meanstructure = TRUE}). To check model convergence, we used the following criteria: (1) the absence of Heywood cases, (2) the maximum absolute element of the gradient vector is less than $10^{-4}$, and (3) the minimum eigenvalue of the Hessian matrix is greater than zero. Default maximum number of iterations was used. In Step 3, the estimated ACM was obtained following the process described in Section 2.2.2. with $M=10,000$. The inverse of the expected information matrix was obtained using the {\tt inverted.information} option in the {\tt lavInspect} function, and the gradient of the individual log-likelihood, $\nabla_{\boldsymbol{\xi}} \log f(\bY_i; \hat{\boldsymbol{\xi}})$, within the formula $\mathbf{A}$ (Equation \ref{eq:ACM1}) was obtained from the {\tt lavScores} function. In Step 4, $19$ evenly spaced values between $-3$ and $3$ were used as evaluation points per LV, resulting in a total of $19^2=361$ pointwise statistics (i.e., $Q=361$) for evaluating the fit at every possible combination of LVs. For the summary statistic, a subgrid of $7^2=49$ points, obtained from 7 equally spaced values from $-2$ to $2$ per LV, were used for illustration purposes. In addition, we set the degree of freedom $s=1$ in Equation \ref{eq:summary} so that the reference limiting distribution for $T_1$ becomes $\chi_1^2$.

\subsubsection{3.1.2. Results}
\begin{figure}[t]
    \centering
    \includegraphics[width=130mm]{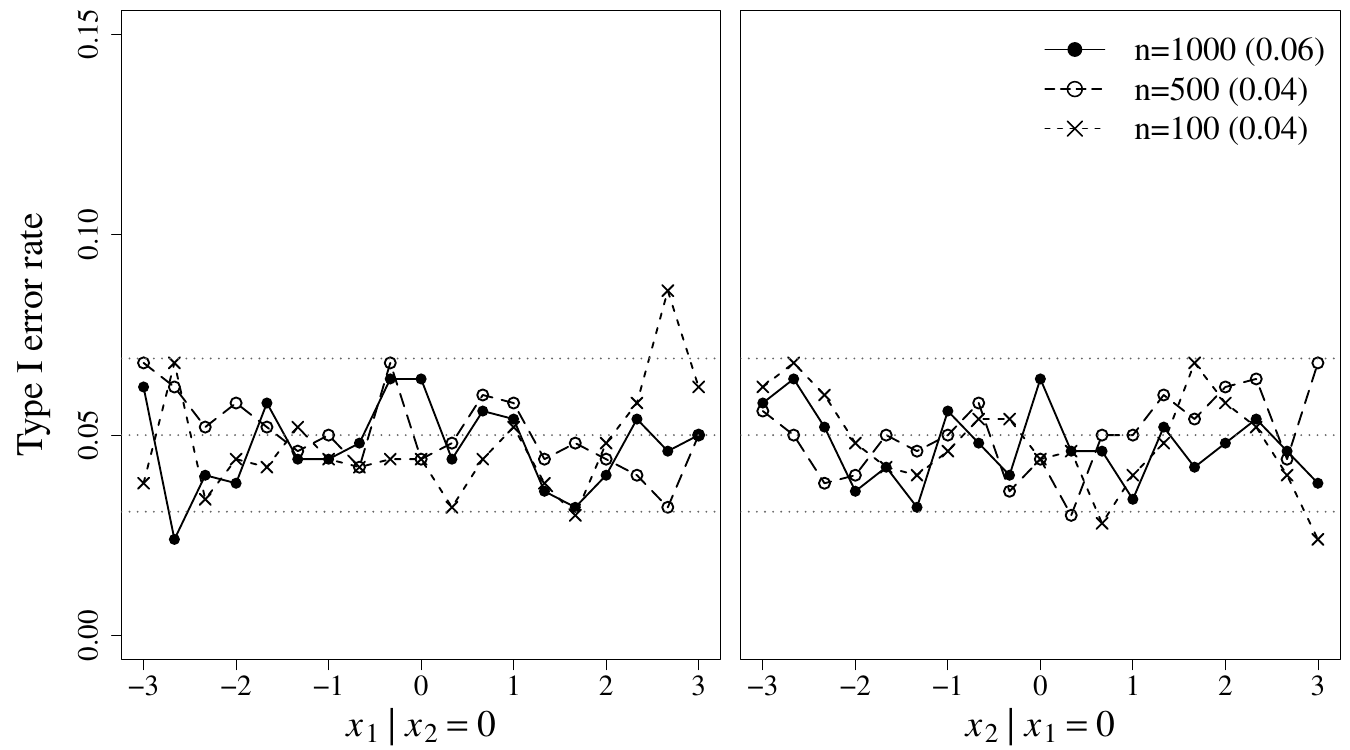}
    \caption{Type I error results for the pointwise and overall LV density fit tests at $\alpha=0.05$. Grey-scaled horizontal dotted lines represent the nominal level $\alpha=0.05$ and the Monte Carlo confidence band. Pointwise $z$-test results are summarized for each LV, conditional on the mean of the other LV. Overall $\chi^2$-test results are presented in parentheses next to the sample size legend.}
    \label{fig:type1}
\end{figure}

% Type I error rate
Empirical Type I error rates were investigated under the correctly specified condition at the nominal level of $\alpha=0.05$. If the proposed tests perform well, the Type I error rates should closely match the nominal level. Figure \ref{fig:type1} depicts the Type I error results for the pointwise $z$-tests (presented as points connected by lines) and the overall $\chi^2$-tests (presented as numbers in parentheses alongside the sample sizes) under various sample size conditions. The left and right panels of Figure \ref{fig:type1} show the results evaluated at each $x_1$ and $x_2$ value, respectively, conditional on the mean of the other LV, which is zero. The results were summarized this way to avoid the influence of LV combinations that are highly unlikely to occur, such as $\bx=(-3, 3)^\top$. In addition, 95\% normal-approximation confidence intervals were obtained around the nominal $\alpha$-level, which indicate an acceptable range of Type I error rates considering Monte Carlo error.

The results indicate that Type I error rates for both pointwise and overall tests are well-controlled across all sample size conditions. For pointwise tests, most of the points in Figure \ref{fig:type1} fall within the Monte Carlo confidence band, suggesting well controlled Type I error rates. Although a few points fall noticeably outside the confidence band at extreme LV values, these extreme values have a low chance of occurrence in practice given the normal density, and the extent of their deviations is also minimal. For the summary results, the Type I error rates closely align with the nominal level across all sample size conditions, indicating satisfactory performance of $T_1$.

\begin{figure}[t]
    \centering
    \includegraphics[width=130mm]{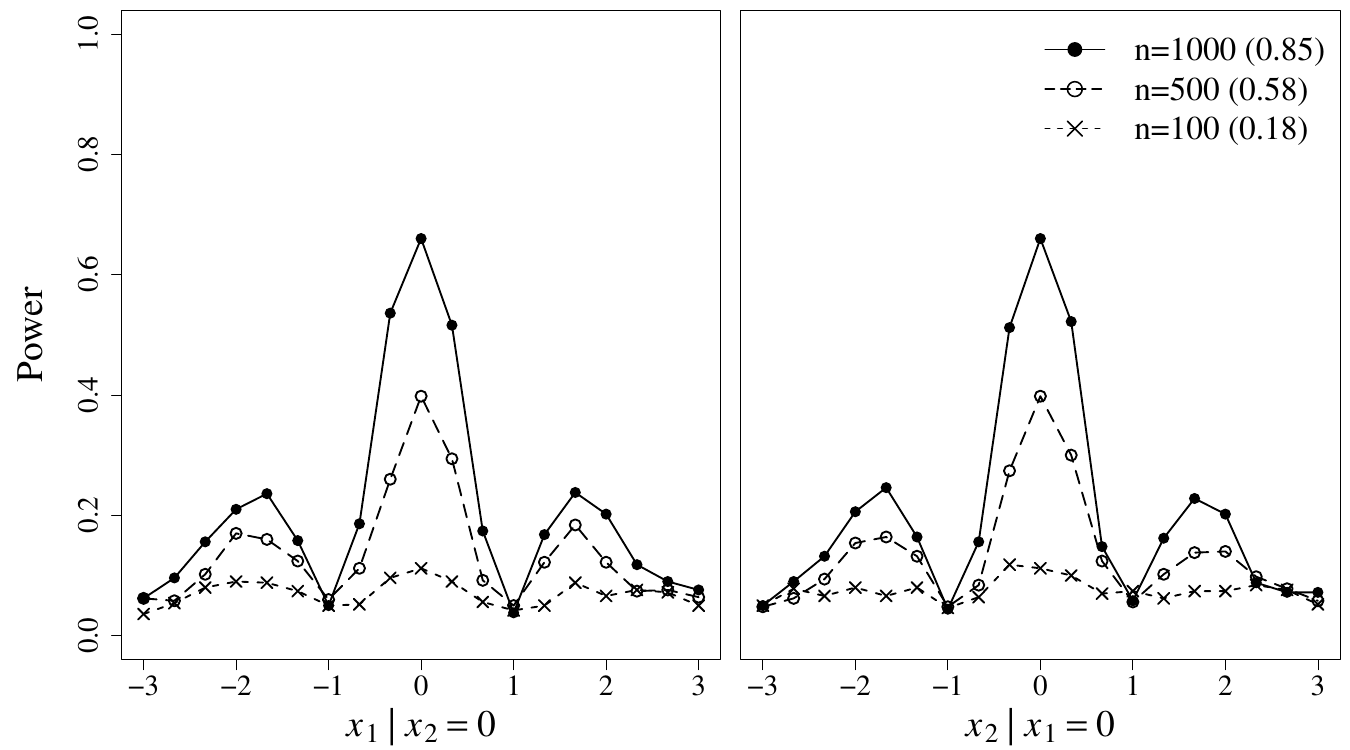}
    \caption{Power results for the pointwise and overall LV density fit tests. Pointwise $z$-test results are summarized for each LV, conditional on the mean of the other LV. Overall $\chi^2$-test results are presented in parentheses next to the sample size legend.}
    \label{fig:power}
\end{figure}

% Power
Empirical power was investigated next under the misspecified condition. Figure \ref{fig:power} displays the power results for the pointwise $z$-tests and the overall $\chi^2$-tests under various sample size conditions. The results were summarized in the same way as in Figure \ref{fig:type1}. 
In general, both pointwise and overall tests demonstrate increasing power in detecting misfit as the sample size increases. While the tests show limited power with a small sample size ($n=100$), they exhibit satisfactory power with moderate to large sample sizes ($n=500$ and $n=1000$). 

Specifically, the power results for the pointwise $z$-tests are consistent  with the shape of the data-generating LV densities shown in the right panel of Figure \ref{fig:LVD}. When constructing the pointwise statistics, recall that $\hat\eta_\ell(\hat\xxi)$ attempts to recover the LV density under the misspecified condition (depicted with black dashed lines in Figure \ref{fig:LVD}), while $\eta_\ell(\hat\xxi)$ is simply a unimodal normal density in this case (depicted with grey solid lines\footnote{The grey solid lines in Figure \ref{fig:LVD} correspond to the values of $\eta_\ell(\xxi)$, whereas the quantity actually compared in $z_1(\bx_\ell)$ is $\eta_\ell(\hat\xxi)$, where $\hat\xxi$ is obtained under misspecification. In this simulation setting, $\eta_\ell(\hat\xxi)$ under misspecification approaches the bivariate normal density $\eta_\ell(\xxi)$ as $n \to \infty$, because the covariance matrices in both conditions are the same in our data-generating model (Equation \ref{eq:phi}) and ML estimation consistently recovers the covariance structure.}). The values of $z_1(\bx_\ell)$ are derived from the discrepancy between these two; therefore, power is expected to be high in regions with large discrepancies and relatively low near crossover points. In our simulation study, both LVs responded to these varying degrees of discrepancy, as indicated by the oscillating pattern of results in Figure \ref{fig:power}. 

\begin{table}[ht]
\caption{Results from the conventional fit diagnostics under the misspecified condition. Satorra-Bentler correction was applied to all statistics. The second column shows rejection rates of the Satorra-Bentler adjusted $\chi^2$ model fit tests. The third to sixth columns present average values of the fit indices over 500 replications. Standard errors are presented in parentheses.}
\begin{center}
\begin{tabular}{c c c c c c}  
\hline
 n  & $\chi^2$ test & CFI & TLI & SRMR & RMSEA\\
\hline
 $100$ & 0.26 (0.020) & 0.98 (0.000) & 0.98 (0.001) & 0.06 (0.000) & 0.03 (0.000) \\
 $500$ & 0.07 (0.011) & 1.00 (0.000) & 1.00 (0.000) & 0.03 (0.000) & 0.01 (0.000) \\
 $1000$ & 0.06 (0.011) & 1.00 (0.000) & 1.00 (0.000) & 0.02 (0.000) & 0.00 (0.000) \\
  \hline
\end{tabular}
\end{center}
\label{tb:SB1}
\end{table}

As suggested by a reviewer, we report in Table \ref{tb:SB1} the performance of the Satorra-Bentler adjusted fit diagnostics (Satorra \& Bentler, 2001) under our misspecified condition. The Satorra-Bentler adjustment aims to correct for the effect of non-normal data on the conventional fit measures. Although not presented here, the adjustment resulted in model fit diagnostics similar to those without the correction. In addition, as can be seen in Table \ref{tb:SB1}, the power of the $\chi^2$ model fit test decreased with increasing sample size, converging to the nominal $\alpha$ level. All fit indices investigated (CFI, TLI, SRMR, and RMSEA) also did not respond to the misspecified LV density. These results show that conventional GOF statistics capture only imperfect recovery of mean and covariance structures but not other important aspects of misfit.

Assessing fit based solely on the recovery of mean and covariance structures may be sufficient when examining parameter estimates such as factor loadings. However, when the focus shifts to factors themselves or their relationships, assessing normality at the latent level becomes crucial. In the Study 1 setup, interpreting the correlation coefficient of 0.2 as indicating a positive linear relationship between the two LVs under the assumption of bivariate normality would be highly misleading, as there actually exist two subgroups with negatively correlated LVs. In general, the inter-LV correlation should be interpreted with caution when there is potential violation of normality at the LV level. 

\subsection{3.2. Simulation Study 2}

 \subsubsection{3.2.1. Simulation Setup}
 
In Study 2, the linear normal one-factor model with $X_i \sim \mathcal{N}(0, 1)$ was under investigation with or without the misspecification in MV-level mean/variance functions. Let $\mu_j$ and $\sigma_j^2$ denote the mean and variance of the conditional distribution $Y_{ij}|X_i$. Under the linear normal factor model, $Y_{ij}|X_i=x \sim \mathcal{N}(\mu_j(x), \sigma_j^2(x))$ with the mean function $\mu_j(x)=\nu_j+\lambda_j x$ and variance function $\sigma_j^2(x)=\theta_j$. MV-level fit assessments were performed by investigating the linearity of $\mu_j$ and homoscedasticity of $\sigma_j^2$.

\begin{figure}[ht]
    \centering
    \includegraphics[width=130mm]{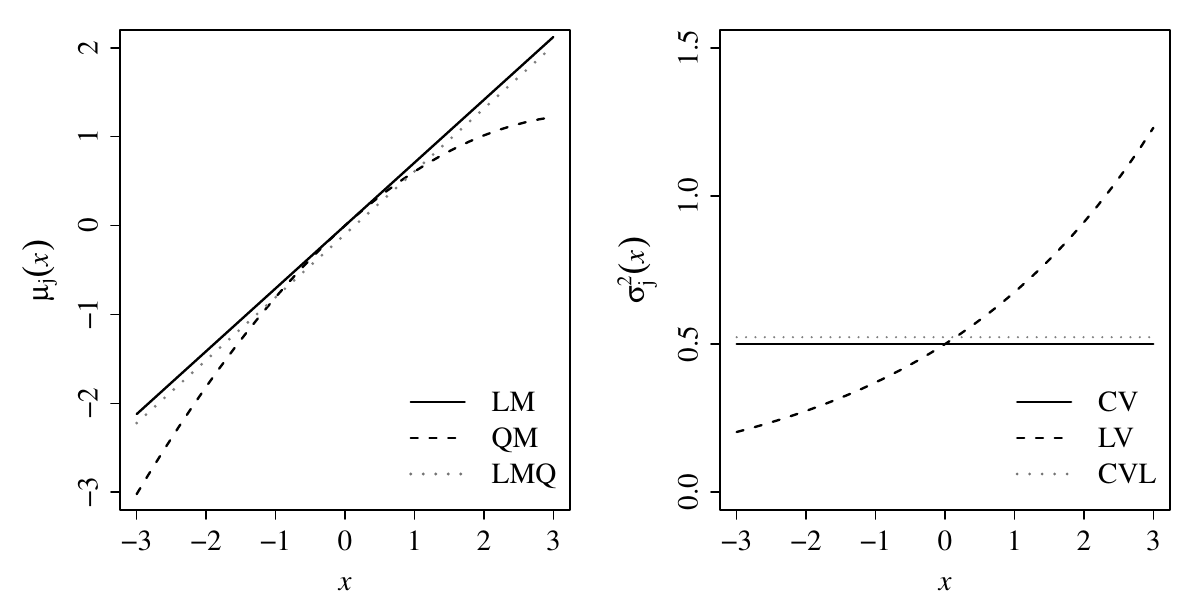}
    \caption{Black lines illustrate the mean function $\mu_j(x)$ (left panel) and variance function $\sigma_j^2(x)$ (right panel) under correctly specified (LM, CV) and misspecified (QM, LV) conditions in the data-generating model. Grey dotted lines (LMQ, CVL) approximate the expected functions under misspecification: a misfitting linear mean under a quadratic population (left) and constant variance under a log-linear population (right). LM: linear mean, QM: quadratic mean, CV: constant variance, LV: log-linear variance, LMQ: linear mean under quadratic population, CVL: constant variance under log-linear population.}
    \label{fig:std2}
\end{figure}

The data-generating model consisted of four types of MVs summarized in Table \ref{tb:MVtypes}. The four types were determined by fully crossing linear vs. quadratic mean functions and constant vs. log-linear variance functions, labeled as LMCV, QMCV, LMLV, and QMLV, respectively. Regardless of the type of MVs, the intercept was set to zero, i.e., $\nu_j=0$, and the error variance was obtained by $\theta_j=1-\lambda_j^2$. The factor loading $\lambda_j$ was set to alternate between $\sqrt{0.3}, \sqrt{0.5},$ and $\sqrt{0.7}$ for LMCVs; $\lambda_j$ was fixed at $\sqrt{0.5}$ for the other three types of MVs. After that, coefficients $\kappa_j$, $\gamma_{0j}$, and $\gamma_{1j}$ were added for the misspecified MVs. The quadratic coefficient $\kappa_j$ for QMCV and QMLV was fixed at $-0.1$; $\gamma_{0j}$ and $\gamma_{1j}$ for LMLV and QMLV were set as $\gamma_{0j}=\log \theta_j$ and $\gamma_{1j}=0.3$, respectively. In Figure \ref{fig:std2}, the black dashed and solid lines illustrate the shapes of $\mu_j(x)$ and $\sigma_j^2(x)$ with and without these additional coefficients, respectively, the combination of which comprises the four types of MVs. As an illustration, the values of $\nu_j=0$, $\lambda_j=\sqrt{0.5}$, and $\theta_j=0.5$ were used in generating the figure.

\begin{table}[t]
\caption{Four types of MVs used for data generation. LMCV: linear mean function, constant variance function. QMCV: quadratic mean function, constant variance function. LMLV: linear mean function, log-linear variance function. QMLV: quadratic mean function, log-linear variance function.}
\begin{center}
\begin{tabular}{l l l l}  
\hline
   Type & Label & Conditional distribution & Misspecification\\
  \hline
  1 & LMCV & $\mathcal{N}(\nu_j+\lambda_j x, \theta_j)$ & Correctly specified\\
  2 & QMCV & $\mathcal{N}(\nu_j+\lambda_j x + \kappa_j x^2, \theta_j)$  & Misspecified mean\\
  3 & LMLV & $\mathcal{N}(\nu_j+\lambda_j x, exp(\gamma_{0j}+\gamma_{1j} x))$  & Misspecified variance\\
  4 & QMLV & $\mathcal{N}(\nu_j+\lambda_j x + \kappa_j x^2, exp(\gamma_{0j}+\gamma_{1j} x))$  & Misspecified mean and variance\\
\hline
\end{tabular}
\end{center}
\label{tb:MVtypes}
\end{table}

For both correctly and incorrectly specified conditions, the number of MVs was fixed at ten ($m=10$). In the correctly specified condition, all ten MVs were LMCVs. In the misspecified condition, three out of ten MVs were misspecified---one for QMCV, LMLV, and QMLV, respectively. The first seven MVs in the misspecified condition were set to be correctly specified to allow for the evaluation of the empirical false detection rate---the proportion of cases in which the tests falsely detect these correctly specified MVs as misfitting MVs in the presence of both correctly and incorrectly specified MVs.

Upon generating the data, the proposed MV-level fit assessments were performed using the following procedures: (1) Estimate the model parameters $\xxi$ by ML to get $\hat\xxi$, (2) construct $\boldsymbol{\hat{\eta}}(\hat\xxi)$ and $\boldsymbol{\eta}(\hat\xxi)$, and apply the ratio transformation to obtain residuals by $\boldsymbol{e}_{\boldsymbol{\varphi}}=\boldsymbol{\varphi}(\boldsymbol{\hat{\eta}}(\hat\xxi))-\boldsymbol{\varphi}(\boldsymbol{\eta}(\hat\xxi))$, (3) estimate the ACM of $\boldsymbol{e}_{\boldsymbol{\varphi}}$, (4) compute $z_2(x_\ell), z_3(x_\ell)$ for $\ell=1,\dots,Q$ and $T_2, T_3$ to conduct pointwise $z$-tests and the overall $\chi^2$-tests. Computational details follow those introduced in Section 3.1.1. for Study 1, except for the number of evaluation points used. In Step 4, we adopted $31$ evenly spaced LV values between $-3$ and 3 as evaluation points (i.e., $Q=31$) for $z_2(x_\ell)$ and $z_3(x_\ell)$. Among these, 11 sub-points ranging from $-2$ to $2$ were selected to construct $T_2$ and $T_3$ for illustration. 

\subsubsection{3.2.2. Results}
\begin{figure}[t]
    \centering
    \includegraphics[width=130mm]{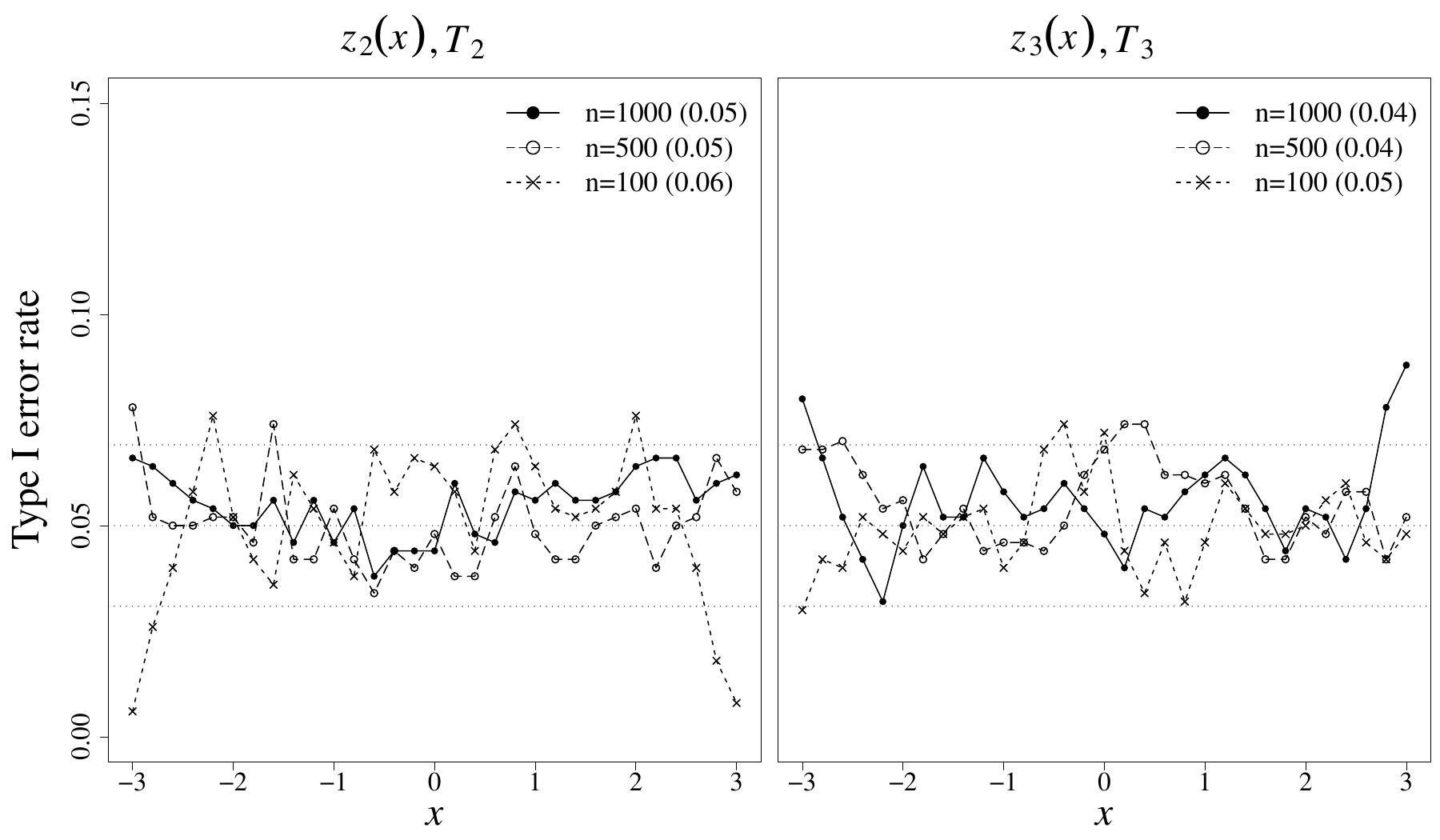}
    \caption{Type I error results for the pointwise and overall MV-level fit tests at $\alpha=0.05$. The left panel shows the results for $z_2(x)$ and $T_2$; the right panel shows the results for $z_3(x)$ and $T_3$. Grey-scaled horizontal dotted lines represent the nominal level $\alpha=0.05$ and the Monte Carlo confidence band. Pointwise $z$-test results are presented as points and connected by lines. Overall $\chi^2$-test results are presented within parentheses next to the sample size legends. The results are presented for one MV ($j=2$) in the correctly specified condition. Similar results were observed for the other nine MVs.}
    \label{fig:type1_2}
\end{figure}

% Type I error rate
Empirical Type I error rates were investigated under the correctly specified condition at the nominal level of $\alpha=0.05$. Figure \ref{fig:type1_2} illustrates the Type I error results for the pointwise $z$-tests (presented as points connected by lines) and the overall $\chi^2$-tests (presented as numbers in parentheses alongside the sample sizes) under various sample size conditions.
The performance of $z_2$ and $T_2$, which assess the linearity of $\mu_j$, is presented in the left panel, while the performance of $z_3$ and $T_3$, which assess the homoscedasticity of $\sigma_j^2$, is presented in the right panel. In the figure, 95\% normal-approximation confidence intervals are delineated around the nominal $\alpha$-level, indicating an acceptable range of Type I error rates considering Monte Carlo error. Results for only one MV ($j=2$ with communality 0.5) are presented due to space limit, but similar results were found for the other nine MVs in the correctly specified condition.

The results demonstrate that Type I error rates are well-controlled for both pointwise and overall tests across all sample size conditions. In the pointwise results, the majority of points in Figure \ref{fig:type1_2} fall within the Monte Carlo confidence band, suggesting decent performance of $z_2$ and $z_3$. Some noticeable deviations are observed in the extreme range of the LV; however, these extreme LV values close to $-3$ and 3 under the standard normal density have a rare chance of occurrence in practice, making them of no significant concern. Overall $\chi^2$-test results closely match with the nominal level across all sample size conditions, verifying the satisfactory performance of $T_2$ and $T_3$.

\begin{figure}[ht!]
    \centering
    \includegraphics[width=165mm]{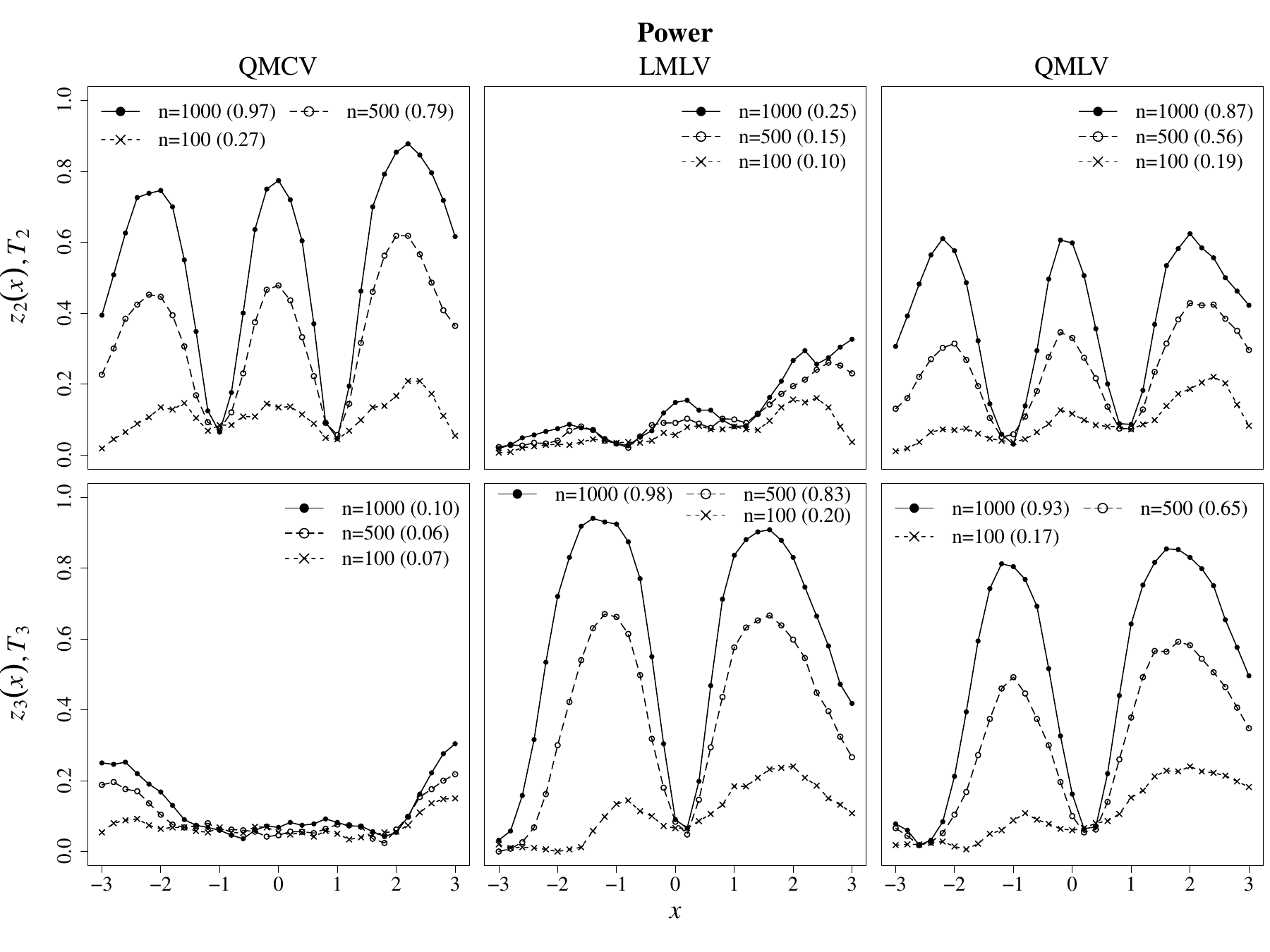}
    \caption{Power results for the pointwise and overall MV-level fit tests. The first row of the figure shows the results for $z_2(x)$ and $T_2$; the second row shows the results for $z_3(x)$ and $T_3$. Each column represents results for QMCV, LMLV, and QMLV (i.e., $j=8, 9,$ and $10$ in the misspecified condition). Pointwise $z$-test results are presented as points and connected by lines. Overall $\chi^2$-test results are presented within parentheses next to the sample size legends.}
    \label{fig:pw}
\end{figure}

% Power
Next, empirical power was investigated under the misspecified condition. The first row of Figure \ref{fig:pw} presents results for $z_2(x)$ and $T_2$, and the second row shows results for $z_3(x)$ and $T_3$. Each column contains results for QMCV, LMLV, and QMLV, respectively.
The proposed MV-level fit tests exhibit distinct behaviors in terms of power depending on the types of MVs. Specifically, $z_2(x)$ and $T_2$ show decent power only against QMCV and QMLV, indicating that they are mainly responsive to detecting misspecification in $\mu_j$ but not in $\sigma_j^2$. On the other hand, $z_3(x)$ and $T_3$ show decent power only against LMLV and QMLV, demonstrating the opposite pattern. These results suggest that the two types of statistics, developed respectively for testing $\mu_j$ and $\sigma_j^2$, perform their intended roles effectively. However, with the small sample of $n=100$, the power tends to be low for all statistics, with little differentiation observed in their respective roles.

Another notable observation is the oscillating pattern in the power of the $z$-tests, which is closely related to the shapes of $\mu_j$ and $\sigma_j^2$ illustrated in Figure \ref{fig:std2}. In this figure, the grey dotted lines (LMQ in the left panel and CVL in the right) represent the expected linear mean and constant variance functions under misspecification, approximated by the average fitting from the $n=1000$ condition in Simulation Study 2. To evaluate the fit, LMQ and CVL are compared to the empirical estimates of QM and LV, respectively. Accordingly, power is expected to be low around points where QM and LMQ (or LV and CVL) intersect and increase as the discrepancy between the two widens. This expectation is confirmed by the simulation results shown in Figure \ref{fig:pw}, where both $z_2(x)$ and $z_3(x)$ exhibit oscillating power patterns that mirror the shape depicted in Figure \ref{fig:std2}, within
approximately the 95\% range of the LV. Notably, the misspecified linear mean function under a quadratic population (LMQ) intersects the true quadratic mean (QM) at two distinct points rather than one, which explains the elevated power of $z_2(x)$ in the middle range.

\begin{figure}[t]
    \centering
    \includegraphics[width=130mm]{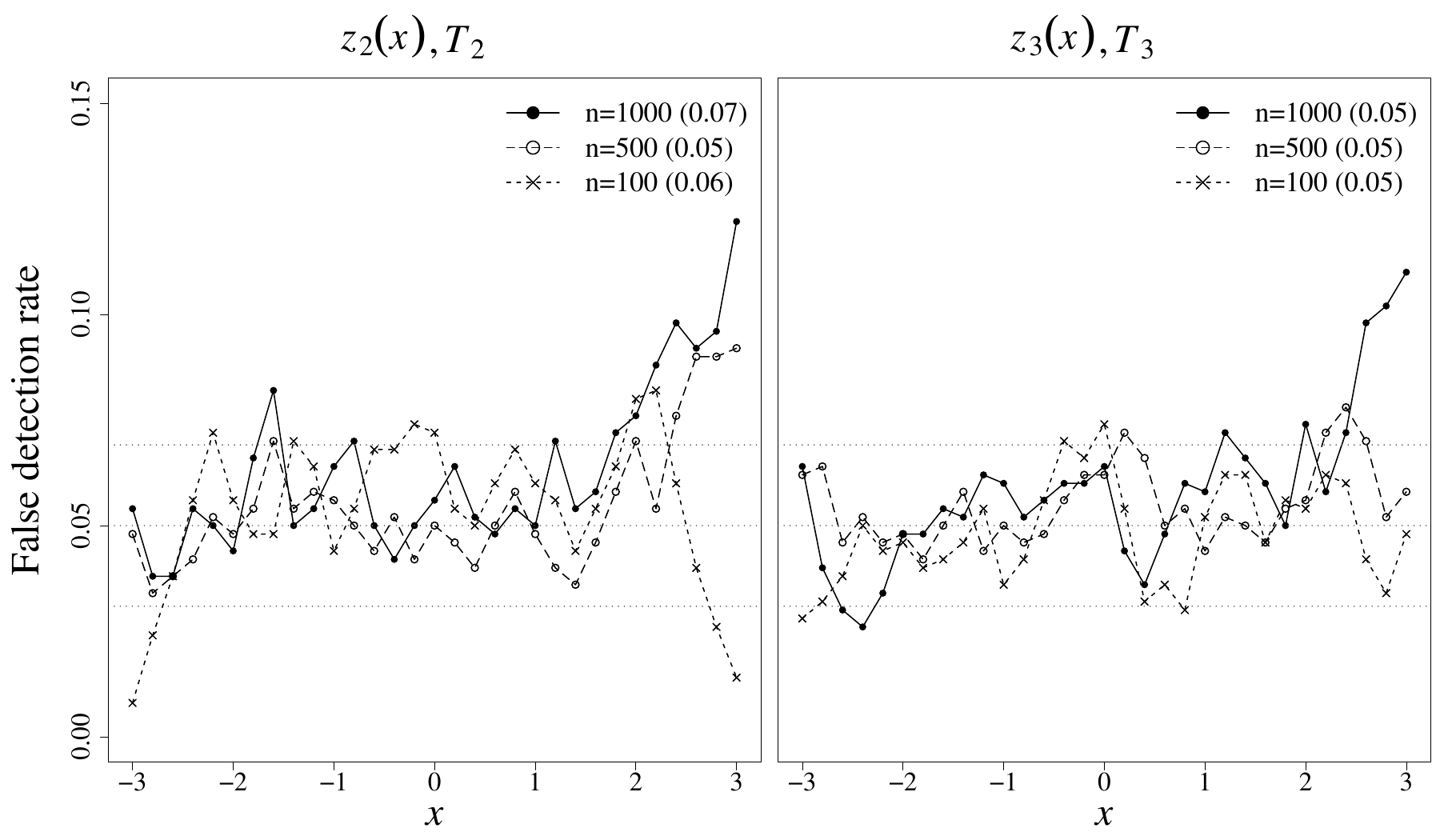}
    \caption{False detection rates for the pointwise and overall MV-level fit tests at $\alpha=0.05$. The left panel shows the result for $z_2(x)$ and $T_2$; the right panel shows the results for $z_3(x)$ and $T_3$. Grey-scaled horizontal dotted lines represent the nominal level $\alpha=0.05$ and the Monte Carlo confidence band. Pointwise $z$-test results are presented as points and connected with lines. Overall $\chi^2$-test results are presented within parentheses next to the sample size legends. The results are presented for one LMCV ($j=2$) in the misspecified condition. Similar results were observed for the other six LMCVs.}
    \label{fig:fdr}
\end{figure}

Finally, empirical false detection rates were examined using correctly specified MVs (i.e., LMCVs) in the misspecified condition. The results are presented in Figure \ref{fig:fdr}, in which the performance of $z_2(x)$ and $T_2$ is presented in the left panel, while the performance of $z_3(x)$ and $T_3$ is presented in the right panel. Again, we provide results for only one MV ($j=2$ with communality 0.5); similar results were observed for the other six LMCVs in the misspecified condition.

In the presence of misspecified MVs in a model, every correctly specified MV will eventually be detected as misfit with a large enough sample size. Hence, the extent to which the rate of false detection is controlled at a low level is another crucial evaluation criterion for MV-level fit tests. In our simulation setup, false detection rates for both pointwise and overall statistics were well-controlled without substantial inflation, indicating their capability of distinguishing between good-fitting and bad-fitting MVs. The rates were similar to the nominal $\alpha$-level in the middle range of the LV. Some inflation was observed in the extreme high end of the LV with the large sample sizes ($n=500 \text{ and/or } 1000$); however, these LV values close to 3 are again have a rare chance of occurrence in reality. 

Similar to Study 1, conventional fit assessment methods with the Satorra-Bentler correction were evaluated for the misspecified condition. Again, there was little difference between adjusted and unadjusted statistics, and as indicated by the results in Table \ref{tb:SB2}, MV-level misspecifications in our setup were not detected by these conventional GOF diagnostics. Note that MV-level misfit cannot be captured by inspecting marginal residual means and variances for each MV either, as these values are always zero for the linear one-factor model, preventing any insights into MV-level misfit conditional on the LV. 

\begin{table}[ht]
\caption{Results from the conventional fit diagnostics under the misspecified condition. Satorra-Bentler correction was applied to all statistics. The second column shows rejection rates of the Satorra-Bentler adjusted $\chi^2$ model fit tests. The third to sixth columns present average values of the fit indices over 500 replications. Standard errors are presented in parentheses.}
\begin{center}
\begin{tabular}{c c c c c c}  
\hline
  n & $\chi^2$ test & CFI & TLI & SRMR & RMSEA\\
\hline
 $100$ & 0.08 (0.012) & 0.99 (0.000) & 0.99 (0.001) & 0.04 (0.000) & 0.03 (0.001) \\
 $500$ & 0.06 (0.011) & 1.00 (0.000) & 1.00 (0.000) & 0.02 (0.000) & 0.01 (0.000) \\
 $1000$ & 0.08 (0.012) & 1.00 (0.000) & 1.00 (0.000) & 0.01 (0.000) & 0.01 (0.000) \\
  \hline
\end{tabular}
\end{center}
\label{tb:SB2}
\end{table}

\section{4. Empirical Example}

In this section, we demonstrate our proposed fit assessment method using item-level response time (RT) data obtained from an online administration of the Test of Relational Reasoning-Junior (TORRjr) among Chinese elementary and middle school students \parencite{zhao2021}. Among the four subscales of TORRjr, we focus on the antinomy subscale, which consists of eight multiple-choice items ($m=8$), with a sample of $n=744$ observations with no missing data. The same RT data were previously analyzed by \textcite{liu2022} using a semiparametric approach. Additional information about TORRjr is available, for example, in \textcite{alexander2016}.

In the RT literature, van der Linden's log-normal RT model \parencite{van2006} stands out as one of the most widely used models for analyzing response times. This model is essentially a linear normal one-factor model fitted to the log-transformed RT. In this context, the MV corresponds to the item-level log-transformed RT, while the LV $X_i$, assumed to follow a standard normal distribution, represents individual's latent processing speed. Lower values of $X_i$ indicate faster processing speed, whereas higher values indicate slower processing speed. Given this interpretation, we will simply refer to the LV as \textit{latent slowness}.

Before proceeding with our fit assessments, we present in Table \ref{tb:SB_real} the results of traditional GOF assessments for the linear-normal one-factor model fitted to the log-transformed RT data. In line with our simulation setup, all diagnostics were adjusted using the Satorra-Bentler correction. The $\chi^2$ model fit test was statistically significant at $\alpha=0.05$, indicating a poor fit of the model ($\chi_{20}^2=60.35, p < .0001$). However, with a sample size of $n=744$, this result would often be ignored in practice due to the tendency of the $\chi^2$ test to suggest poor fit in large samples, even with trivial amount of misfit. Instead, researchers typically rely on descriptive GOF indices in such cases. Here, values for  CFI=0.97, TLI=0.96, SRMR=0.02, and RMSEA=0.05 all fell within acceptable ranges (according to the rules of thumb by, e.g., \cite{huben1999}). Based on these results, researchers would generally conclude that the model shows a decent fit with the data and proceed with subsequent inferences. In what follows, we illustrate how our fit diagnostics can provide deeper insights using the same data and model.

\begin{table}[ht]
\caption{Results from the conventional GOF diagnostics based on TORRjr response time data, with the Satorra-Bentler correction applied to all statistics.}
\begin{center}
\begin{tabular}{c c c c c c c}  
\hline
  n & $\chi_{20}^2$  & CFI & TLI & SRMR & RMSEA\\
\hline
 $744$ & 60.35 & 0.97 & 0.96 & 0.02 & 0.05 \\
   \hline
\end{tabular}
\end{center}
\label{tb:SB_real}
\end{table}

\subsection{4.1. Fit Diagnostics}

We now apply our proposed test statistics to evaluate the normality of the LV density and item-level linearity and homoscedasticity, all of which are fundamental assumptions in the linear normal factor model. For each analysis described below, 31 equally spaced LV values from $-3$ to 3 were selected as evaluation points, and all of them were used to construct the summary statistic. This summary information can be useful in practical applications where researchers seek an overall judgment based on multiple pointwise assessments. All other specific options were applied consistently with those used in the simulation studies. 

\begin{figure}[t]
    \centering
    \includegraphics[width=130mm]{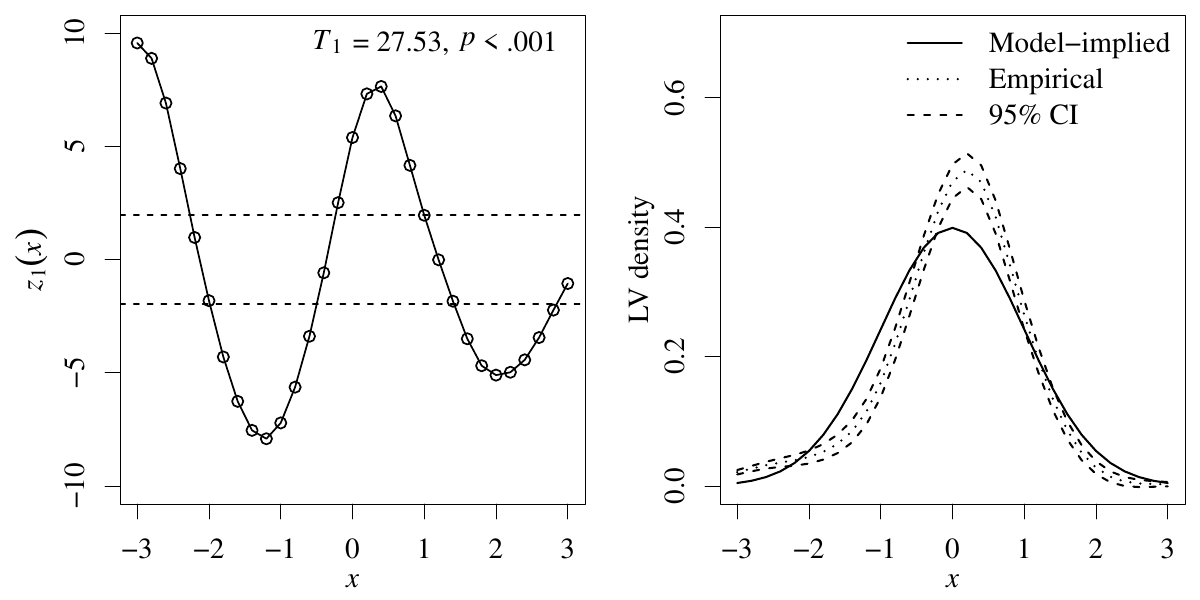}
    \caption{LV density fit test results at $\alpha=0.05$. The left panel displays results with respect to the test statistics $z_1(x)$ and $T_1$. The right panel presents the same information in a more intuitive way. In both panels, dashed lines indicate the 95\% pointwise confidence band; points or segments of the solid line falling outside this band indicate significant misfit.}
    \label{fig:real_LD_new}
\end{figure}

The normality of the LV density was tested using the pointwise statistics $z_1(x_\ell)$, $\ell=1, \dots, 31$, along with the summary statistic $T_1$. Results are summarized graphically in Figure \ref{fig:real_LD_new}. In the left panel, circled points represent the $z_1(x_\ell)$ values, with $T_1$ and the corresponding \textit{p}-value for the overall $\chi^2$-test presented at the top. Horizontal dashed lines at $\pm 1.96$ represent the 95\% pointwise confidence intervals at $\alpha=0.05$. Points falling outside this range (i.e., $|z_1|>1.96$) indicate significant misfit in the LV density. The right panel offers a more intuitive graphical representation of the same information. Here, the solid line depicts the model-implied standard normal density, drawn by connecting the points $\eta_\ell(\hat\xxi)=\phi(x_\ell)$. Similarly, the dotted line represents the empirically estimated LV density, $\hat{\eta}_\ell(\hat{\xxi})$. Dashed lines indicate the 95\% pointwise confidence intervals, calculated by $\hat{\eta}_\ell(\hat{\xxi}) \pm 1.96 [\hat{\sigma}_{\boldsymbol{\varphi},\ell} / \sqrt{n}]$ (from Equation \ref{eq:z} with $\boldsymbol{\varphi}=\boldsymbol{\iota}$). In this plot, the part of the solid line that falls outside the confidence band indicates significant misfit.

As shown in Figure \ref{fig:real_LD_new}, our method identified some violations in the normality assumption of the LV density in this example. The graphical plot in the right panel further reveals that the LV density, as estimated from the data, is leptokurtic and slightly negatively skewed. This finding mirrors the descriptive statistics reported in Table 2 of \textcite{liu2022}, which showed that the observed log-RT distributions were leptokurtic and/or skewed for some items. Practically, this deviation from the standard normal density suggests, for example, that there are substantially more fast responders and fewer slow responders than assumed by the model. For an overall judgment on the LV density fit based on this graphical diagnostic, one can refer to the overall test result, which indicates significant misfit in the LV density in general ($T_1=27.53, \textit{p}<.001$).

\begin{figure}[t]
    \centering
    \includegraphics[width=165mm]{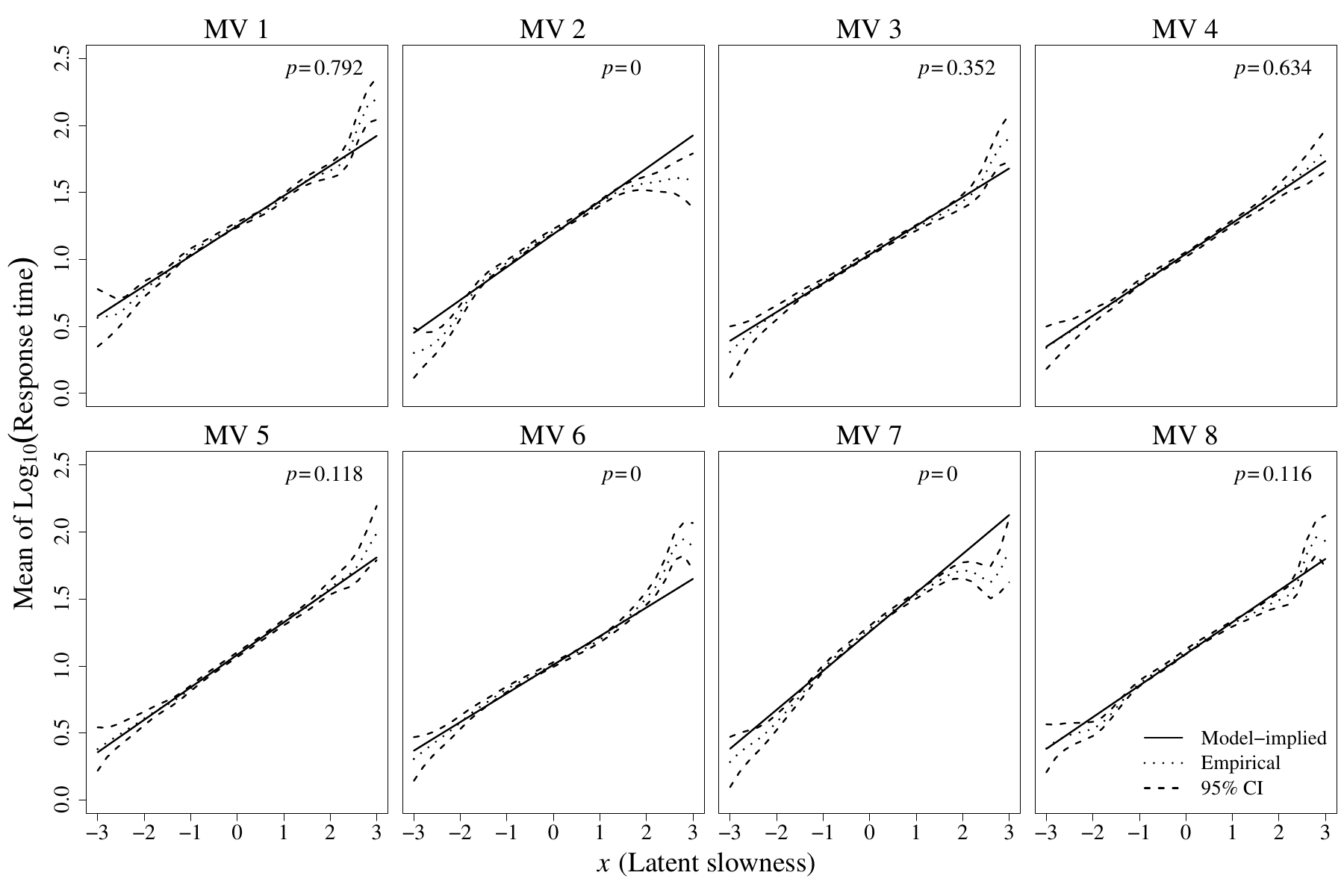}
    \caption{MV-level linearity test results at $\alpha=0.05$. Solid lines represent model-implied linear mean functions, while dotted lines indicate empirical estimates of the mean functions. Dashed lines delineate 95\% pointwise confidence bands. The \textit{p}-value of the overall $\chi^2$-test based on $T_2$ is denoted as \textit{p} in this figure.}
    \label{fig:real_MV1}
\end{figure}

\begin{figure}[t]
    \centering
    \includegraphics[width=165mm]{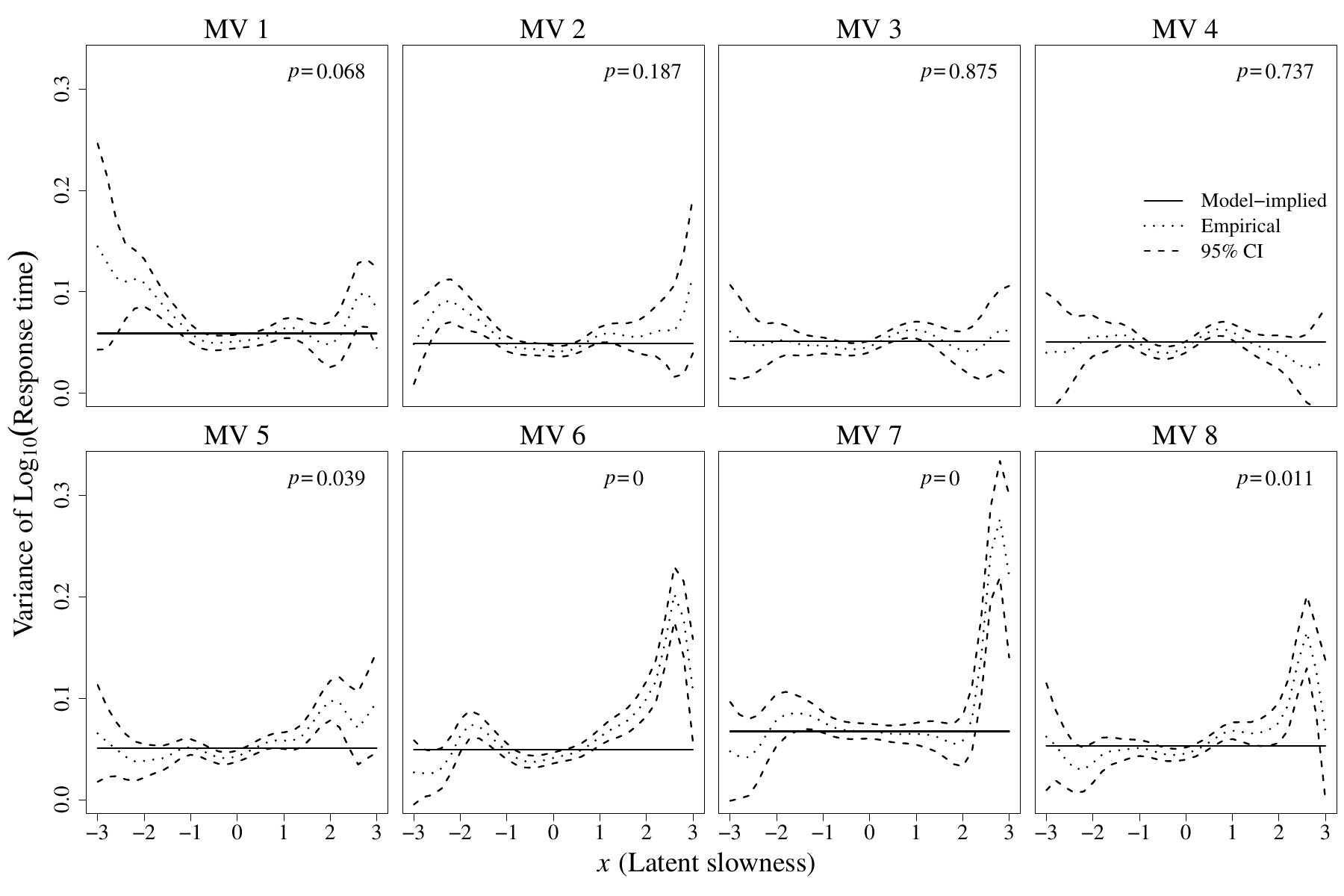}
    \caption{MV-level homoscedasticity test results at $\alpha=0.05$. Solid lines represent model-implied constant variances, while dotted lines indicate empirical estimates of the variance functions. Dashed lines delineate 95\% pointwise confidence bands. The \textit{p}-value of the overall $\chi^2$-test based on $T_3$ is denoted as \textit{p} in this figure.}
    \label{fig:real_MV2}
\end{figure}

Next, item-level fit was evaluated by examining the mean function of the log-RT using $z_2(x)$ and $T_2$, and by examining the variance function using $z_3(x)$ and $T_3$. Figures \ref{fig:real_MV1} and \ref{fig:real_MV2} present the corresponding results for the mean and variance functions, respectively, adopting the second version of the graphical representation shown in Figure \ref{fig:real_LD_new}. In each figure, solid lines represent the model-implied linear mean functions and constant variance functions, respectively, while dotted lines represent their empirical estimates derived from the data. Dashed lines indicate the 95\% pointwise confidence intervals (Equation \ref{eq:z} with $\boldsymbol{\varphi}$ being the ratio transformation). For each item, the \textit{p}-values for the overall $\chi^2$-test based on $T_2 \text{ or } T_3$ are provided at the top of the figures. 

Examining the pointwise results in Figure \ref{fig:real_MV1} reveals that certain items, particularly items 2, 6, and 7, exhibit substantial violations of linearity in their mean functions, as indicated by curvature in their confidence bands. Specifically, item 2 shows a banding-down shape at both extremes of the LVs, suggesting that the misfitting linear normal factor model significantly overestimates the conditional mean of log-RT in these regions. Such misfit can be especially problematic if researchers' focus is on investigating extremely slow or fast responders, which is often the case in RT data analysis. For example, when identifying careless responders based on unusually fast responses, the model's overestimation of the expected RT for fast responders could inflate the false positive rate. Similar discussions can also be found in \textcite{liu2022}, where similar curvature patterns in conditional mean functions have been identified using a semiparametric approach. The overall $\chi^2$-test results presented at the top of each plot in Figure \ref{fig:real_MV1} can be referenced to identify items whose mean functions are poorly recovered by a linear normal factor model. Specifically, items 2, 6, and 7 are called out by the overall tests, offering a formal justification for what might otherwise depend on subjective judgment.

Similarly, the fit assessment of item-level variance functions in Figure \ref{fig:real_MV2} indicates that the constant variance assumption is violated for several items, as evidenced by confidence bands deviating noticeably from the model-implied constant variances. These deviations suggest heteroscedastic variance depending on the level of latent processing speed. Particularly, the salient increase in variability at the high end of the LV for items 6 and 7 reflects greater RT variability among slow responders. Summarizing the pointwise results, the overall tests identified items 5, 6, 7, and 8 as misfitting in this case.

\section{5. Discussion}

To achieve more flexible fit assessment, we extend the theory of generalized residuals \parencite{haberman2013gr} to accommodate more general measurement models. In our framework, transformed residuals are defined to construct asymptotically normal and chi-squared fit statistics, enabling the evaluation of both individual and multiple residuals simultaneously. As specific examples, we propose pointwise and summary statistics to identify misfit in the LV density and MV-level conditional moments under common factor models. Results from two simulation studies indicate that the proposed tests, based on these example statistics, maintain well-controlled Type I error rates even with small sample sizes and exhibit decent power to detect misfit with moderate to large sample sizes. These findings suggest that generalized residuals are promising tools for assessing parametric assumptions involved in measurement models. 

The empirical data analysis using response time data further illustrates the practical utility of our approach. While conventional GOF diagnostics suggest an overall reasonable fit for the linear normal one-factor model, our method identifies misfit in the LV density, the MV-level mean function, and the MV-level variance function. Moreover, we provide intuitive visualizations contrasting empirical versus model-implied estimates of relevant summary quantities conditional on LV values (e.g., Figures \ref{fig:real_LD_new}, \ref{fig:real_MV1}, and \ref{fig:real_MV2}), which facilitates the interpretation of misfitting patterns.

Although not addressed in the current manuscript, diagnostics generated from the proposed method could inform model modification. For instance, if a bimodal LV distribution is suggested in the graphical summary of generalized residuals, one may proceed with mixture modeling to explore the presence of hidden subgroups. In longitudinal measurement models for time-varying constructs, the violation of multivariate normality could indicate a nonlinear growth pattern. Similarly, the detected shape of conditional moment functions for MVs could suggest fitting a nonlinear factor model. In addition, identifying misfitting MVs itself could contribute to establishing test fairness.

There are several limitations in our current work that remain to be addressed in future research. 

First, the proposed framework relies on ML estimation, which requires a correctly specified likelihood. This implies that when the framework is adapted to test a specific assumption, such as linearity, it assumes that all other model assumptions are satisfied except for the one under investigation. As noted in Section 2.4.2. and the supplementary document, the ratio-form estimator developed for MV-level fit tests may perform well even when the LV density is misspecified; however, its robustness against other types of misspecification has not been evaluated. If other aspects of the model are misspecified alongside the one under investigation, the performance of the statistics under our framework could be affected. Future research could explore the influence of mixed misspecifications on generalized residuals through simulation studies. Furthermore, incorporating alternative estimation methods beyond ML could be considered to expand the applicability of our framework.  

Second, the tests developed in our framework share a limitation with the conventional model fit $\chi^2$ test in that they would reject the null hypothesis even under trivial misfit in large samples. Although not considered in the current work, introducing effect size measures, such as those analogous to RMSEA or Cohen's effect sizes \parencite{cohen1988}, could be explored in future work as a potential solution. For instance, Cohen's $d$ and $f$ remove the influence of sample size from the associated $t$- and $F$-tests, and RMSEA rescales the $\chi^2$ fit statistic by both the sample size and the degrees of freedom. Standardized effect size measures based on the proposed $z$- and $\chi^2$- statistics could be developed in a similar fashion.

Third, beyond the three example test statistics provided, other statistics can be constructed under our framework to evaluate different types of model assumptions. One simple application is to formulate generalized residuals regarding the conditional covariance between two MVs, which can yield GOF test statistic for assessing the pairwise local independence assumption (e.g., \cite{ip2001}; \cite{mcdonald1994}; \cite{strout1990}). 

Fourth, although we focused only the common factor model in this paper, the theory we present can be extended to other measurement models, as long as a fully specified parametric model is assumed. For example, the framework can be generalized to GOF assessments in non-linear factor analysis \parencite[e.g.,][]{yalcin2001} and count data IRT (e.g., \cite{wang2010}). 

Finally, exploring efficient ways to compute and visualize generalized residuals in high-dimensional LV settings can be a potential topic for future research. When the latent dimensionality is high, constructing evaluation points using an outer product grid becomes infeasible due to the exponential growth in the number of points. For example, in a five-dimensional case, examining residuals at just five different points per LV leads to $5^5=3,125$ evaluations, making the process computationally intensive and posing a challenge for generating graphical plots. Selecting appropriate grids of LVs to summarize results adds another layer of complexity. Formulating partially marginalized residuals conditioned only on one or two LV(s) may yield more efficient and informative diagnostics.

\printbibliography
\includepdf[pages=-]{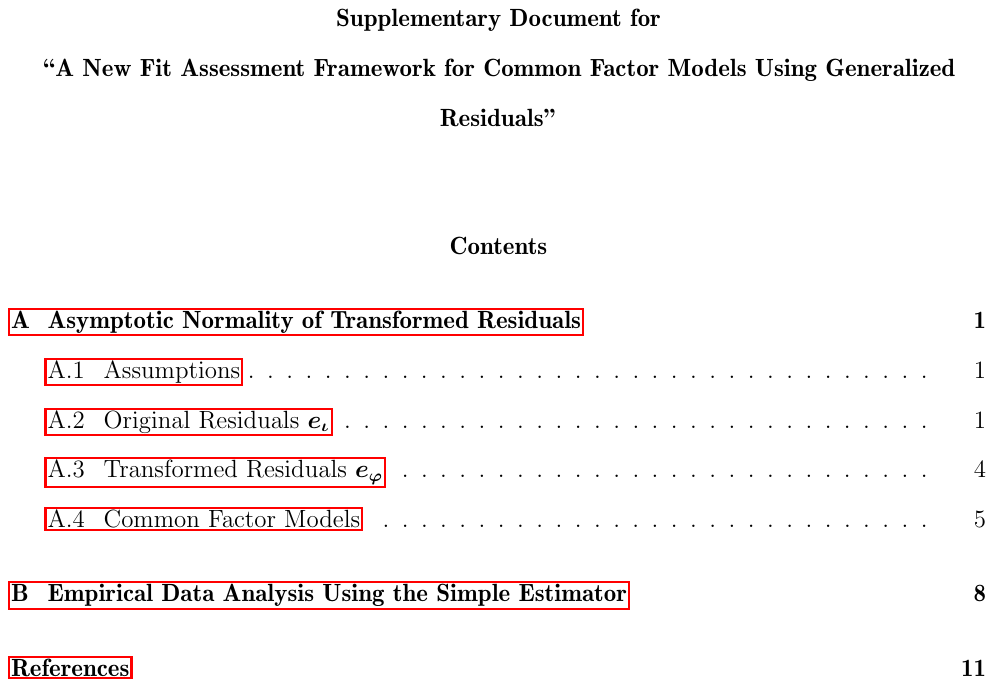}

\end{document}